\documentclass[12pt, draftclsnofoot, onecolumn]{IEEEtran}

\usepackage{cite}
\usepackage{mathtools}
\usepackage{hyperref}
\usepackage{graphicx}
\usepackage{enumitem}
\usepackage{tikz}
\usepackage{bm}
\usepackage{amsmath}
\usepackage{algorithm}
\usepackage{algorithmic}
\usepackage{amsmath,amssymb,amsthm}
\usepackage{array}
\usepackage{relsize}
\usepackage{subcaption}
\usepackage{mathtools, nccmath}
\usepackage[labelformat=simple]{subcaption}
\usepackage[normalem]{ulem}
\newcolumntype{P}[1]{>{\centering\arraybackslash}p{#1}}
\DeclareMathOperator{\C}{\mathbb{C}}
\DeclareMathOperator{\E}{\mathbb{E}}
\DeclareMathOperator{\R}{\mathbb{R}}
\DeclareMathOperator{\N}{\mathcal{N}}
\DeclareMathOperator{\M}{\mathcal{M}}
\DeclareMathOperator{\LA}{\mathcal{L}}
\DeclareMathOperator{\CG}{\mathcal{C}}
\DeclareMathOperator{\BO}{\mathcal{O}}

\DeclareMathOperator*{\argmin}{arg\,min}
\DeclarePairedDelimiter\abs{\lvert}{\rvert}
\DeclareCaptionSubType * [alph]{table}
\newcommand{\x}{\mathbf{X}}
\newcommand{\A}{\mathbf{A}}
\newcommand{\B}{\mathbf{B}}
\newcommand{\CC}{\mathbf{C}}
\newcommand{\h}{\mathbf{H}}
\newcommand{\f}{\mathbf{F}}

\newcommand{\y}{\mathbf{Y}}

\newcommand{\I}{\mathbf{I}}
\newcommand{\z}{\mathbf{Z}}
\newcommand{\g}{\mathbf{G}}
\newcommand{\e}{\mathbf{E}}
\newcommand{\K}{\mathbf{K}}
\newcommand{\U}{\mathbf{U}}
\newcommand{\V}{\mathbf{V}}
\newcommand{\m}{\mathbf{M}}

\newcommand{\rt}{\mathbf{R}}
\newcommand{\D}{\mathbf{D}}
\newcommand{\as}{\mathbf{a}}
\newcommand{\bs}{\mathbf{b}}

\newcommand{\js}{{\rm j}}
\newcommand{\cn}{\mathsf{c}}

\newcommand{\vs}{\mathbf{v}}
\newcommand{\us}{\mathbf{u}}

\newcommand{\hs}{\mathbf{h}}

\newcommand{\xs}{\mathbf{x}}
\newcommand{\xsn}{\mathsf{x}}
\newcommand{\xxn}{\bm{\mathsf{x}}}
\newcommand{\xn}{\bm{\mathsf{X}}}
\newcommand{\ysn}{\mathsf{y}}
\newcommand{\s}{\mathbf{S}}
\newcommand{\ssb}{\mathbf{s}}
\newcommand{\sn}{\bm{\mathsf{S}}}
\newcommand{\ssn}{\bm{\mathsf{s}}}
\newcommand{\ssf}{\mathsf{s}}
\newcommand{\yn}{\bm{\mathsf{Y}}}
\newcommand{\ys}{\bm{\mathsf{y}}}
\newcommand{\ws}{\bm{\mathsf{w}}}
\newcommand{\wn}{\bm{\mathsf{W}}}

\newcommand{\zs}{\mathbf{z}}

\newcommand{\hr}{\mathsf{H}}
\newcommand{\tr}{\mathsf{T}}
\newcommand{\1}{\mathbf{1}}
\newcommand{\0}{\mathbf{0}}
\newcommand{\bc}{\begin{center}}
\newcommand{\ec}{\end{center}}
\newcommand{\norm}[1]{\left\lVert#1\right\rVert}
\newcommand{\ds}{\displaystyle}
\newcommand{\uprightsubscript}[1]{_{\textnormal{#1}}}
\begingroup\lccode`~=`?\lowercase{\endgroup\let~}\uprightsubscript
\AtBeginDocument{\mathcode`?="8000 }
\newcommand{\<}[1]{^{\textnormal{#1}}}
\newcommand{\tn}[1]{\textnormal{#1}}

\theoremstyle{remark}
\newtheorem*{remark}{Remark}
\definecolor{dg}{RGB}{0,0,0}

\begin{document}
\title{Modulating Intelligent Surfaces for Multi-User MIMO Systems: Beamforming and Modulation Design}
\author{\IEEEauthorblockN{Haseeb Ur Rehman, Faouzi Bellili, \textit{Member, IEEE}, Amine Mezghani, \textit{Member, IEEE}, and Ekram Hossain, \textit{Fellow, IEEE}} \thanks{The authors are with the Department of Electrical and Computer Engineering at the University of Manitoba, Canada (emails: urrehmah@myumanitoba.ca, \{Faouzi.Bellili, Amine.Mezghani, Ekram.Hossain\}@umanitoba.ca). This work was supported by the Natural Sciences and Engineering Research Council of Canada (NSERC) and Futurewei Technologies.}}
\date{}
\maketitle

\begin{abstract}
This paper introduces a novel approach of utilizing the reconfigurable intelligent surface (RIS) for joint data modulation and signal beamforming in a multi-user downlink cellular network by leveraging the idea of backscatter communication. We present a general framework in which the RIS, referred to as modulating intelligent surface (MIS) in this paper, is used to: $i)$ beamform the signals for a set of users whose data modulation is already performed by the base station (BS), and at the same time,  $ii)$  embed the data of a different set of users by passively modulating the deliberately sent carrier signals from the BS to the RIS. To maximize each user's spectral efficiency, a joint non-convex optimization problem is formulated under the sum minimum mean-square error (MMSE) criterion. Alternating optimization is used to divide the original joint problem into two tasks of: $i)$ separately optimizing the MIS phase-shifts for passive beamforming along with data embedding for the BS- and MIS-served users, respectively, and $ii)$ jointly optimizing the active precoder and the receive scaling factor for the BS- and MIS-served users, respectively. While the solution to the latter joint problem is found in closed-form using traditional optimization techniques, the  optimal phase-shifts at the MIS are obtained by deriving the appropriate optimization-oriented vector approximate message passing (OOVAMP) algorithm. Moreover, the original joint problem is solved under both ideal and practical constraints on the MIS phase shifts, namely, the unimodular constraint and assuming each MIS element to be terminated by a variable reactive load. The proposed MIS-assisted scheme is compared against state-of-the-art RIS-assisted wireless communication schemes and simulation results reveal that it brings substantial improvements in terms of system throughput while supporting a much higher number of users. 
\end{abstract}


\section{Introduction}


\subsection{Background}
Reconfigurable intelligent surfaces (RISs) have recently emerged as a promising technology for beyond-5G/6G wireless communications as it can improve both the spectral and energy efficiencies of wireless systems. RISs also offer the benefits of low-cost and easy integration into the currently deployed wireless systems \cite{rapirs,wuirs2,huirs,irssurvey}. Specifically, a RIS consists of an array of reconfigurable passive reflective elements which enable one to control the wireless propagation environment. RIS is an energy-efficient technology since it allows  to passively beamform the incoming signals without the need for a power amplifier as required in traditional multiple-input multiple-output (MIMO) base stations (BSs) \cite{ngomimo, larssonmimo,irssurvey,guoris}. It does so by suitably optimizing the phase shifts applied by each reflective element to constructively combine the incoming signals so as to achieve improved received power at the end users. The active transmit precoding at the BS together with the RIS phase shifts (for passive beamforming) can be jointly optimized in order to maximize the spectral-efficiency \cite{haseebirs,wuirs,basarirs,ganris,guoris} or the energy-efficiency of the system \cite{huangirs, youirs}.
Moreover, a RIS can also be utilized to simultaneously perform passive beamforming and physical information transfer \cite{yanirs,linris} (e.g., synchronization data or the channel state information [CSI] estimated at the RIS). In \cite{zhaoirs}, the authors presented a massive backscatter wireless communication (MBWC) scheme to encode information on Wi-Fi signals reflected by the RIS.
So far, {\em the vast majority of the work has focused on utilizing the RIS for passive beamforming and the idea of using it also to embed information in a multi-user cellular network has not been explored}. In this paper, by leveraging the idea of backscatter communication \cite{zhaoirs,xuback,yaoback}, we propose a general framework in which the RIS is used for data embedding on the reflected or re-emitted signals by changing the imdepence of each RIS reflective element. In the proposed framework, the RIS can be used to either: $i)$ perform passive beamforming for users served by a BS, or $ii)$ embed information through backscatter communication, or $iii)$ do both simultaneously. In this setting, the RIS phase shifts vary with the inevitable changes in the propagation medium to perform passive beamforming for one set of users, and also with every transmit symbol vector in order to modulate the carrier by data for another set of users. We name the smart surface with such capability as ``modulating intelligent surface (MIS)" \footnote{Such type of antenna surfaces is also known as space-time-coding digital metasurfaces in the antenna literature \cite{zhang2018}.}.

\subsection{Motivation}
The benefits of MISs, that can beamform and modulate signals  at the same time, are three-fold:
\begin{enumerate}
    \item In traditional purely reflective RIS-based schemes, the users' received signals are subject to  severe attenuations stemming from the product path-loss of the BS-RIS and RIS-user links since both links are used for data communication.  Using a MIS, however, allows the BS to focus all of the transmit power (of a reference signal) towards the strongest path in the BS-MIS link and the modulated signals are passively generated at the MIS by appropriately designing its phase shifts. In this way, the information-bearing signals undergo the path-loss of the MIS-user link only before reaching the intended users.
    \item [ii.] The total number of users which can be served is not limited by the number of channel paths that are available in the BS-MIS link, but rather by the number of reflective elements in the MIS. Therefore, more users than the number of BS antennas can be served by allocating power towards a MIS having a high number of antenna elements and then serving the users through the MIS. 
    \item [iii.] The MIS can serve users by recycling the incoming signals transmitted by the BS without any RF chains, thus making the entire approach very cost-effective.
\end{enumerate}
A practical implementation of the proposed MIS-based scheme  requires the BS in part to: $i)$ convey the optimized phase shifts to the MIS through a high speed communication link, and $ii)$  radiate a carrier RF signal to provide the MIS with the required power for passively embedding the data of a predefined set of users. Ideally, the MIS must be installed at a location that is in  line-of-sight (LOS) to the BS to minimize the power loss. There are two possible ways to transmit the optimized phase shifts to the MIS: $i)$ through an optical fibre link between the BS and the MIS, $ii)$ through a high-speed reliable wireless link (e.g., Terahertz wireless links) between the BS and the MIS, if they are in close proximity to each other.
\subsection{Contributions}
We consider a single-cell downlink MIMO system assisted by  a single MIS which is  equipped  with  a  large  number  of  passive  phase shifter elements. The MIS  helps the BS in better serving one portion of the users through passive beamforming and also embeds the information-bearing data for the remaining users on a separate carrier signal that it receives from the BS. In this regard, we build on our recent work in \cite{haseebirs} and propose a general method to jointly optimize:
\begin{itemize}
    \item the active BS precoder,
    \item the receive scaling factors for the BS- and MIS-served users,
    \item the MIS phase shifts used for passive beamforming towards the BS-served users, and
    \item the  data embedding on the reflected signal (i.e., modulation of the reflected signal) for the MIS-served users.
\end{itemize}
 The major contributions embodied by this paper are summarized as follows.
\begin{itemize}
    \item To the best of our knowledge, this is the first work that studies the use of  RISs  for passive beamforming and data embedding at the same time in a multi-user setup. We solve the problem of maximizing the users' spectral-efficiencies  by jointly optimizing the transmit precoding matrix at the BS, the receive scaling factor for the BS- and MIS-served users, and the MIS phase shifts.
    By building on our recent work in \cite{haseebirs}, we formulate the joint optimization problem under the sum minimum mean-square error (MMSE) criterion in order to minimize the mean-square error (MSE) of the received symbols for all users at the same time.
    
    \item To solve the underlying joint optimization problem, we first split it using alternating optimization \cite{dimitrinl} into two simpler sub-optimization tasks, one for finding the optimum MIS phase shifts and the other for jointly optimizing the active BS precoder and the receive scaling factors for both the BS- and MIS-served users. The solution to the latter sub-optimization task is found in closed form.
    
    \item We apply the OOVAMP algorithm to optimize the MIS phase shifts under two different constraints, namely, $i)$ the unimodular constraint, and $ii)$ assuming each MIS element to be terminated by a tunable reactive load. 
    
    \item We present various numerical results to compare the proposed scheme against the standard approach in which a RIS is used for passive beamforming only while  active MMSE  precoding is used at the BS for all users. In this context, we consider two baseline techniques that rely on $i)$  semi-definite relaxation (SDR) \cite{wusdrc,wuirs}, and $ii)$ OOVAMP-based alternating optimization \cite{haseebirs} to find the adequate RIS phase shifts. The latter approach boils down to a special case of the herein proposed scheme when the number of MIS-served users is equal to zero. The simulation results show that using  MISs for joint beamforming and information embedding significantly outperforms the classical schemes in which RISs are used for passively beamforming the signals they receive from the BS. We also study the resilience of the proposed scheme under  CSI mismatches stemming from imperfect CSI acquisition in practice.
\end{itemize}

\subsection{Paper Organization and Notations}
The rest of this paper is organized as follows. The system model along with the problem formulation for jointly optimizing the active BS precoder and the MIS phase-shifts are discussed in Section \ref{sec:system-model}. Section \ref{sec:vamp-opt} discusses the OOVAMP algorithm and its constituent modules. In Section \ref{sec:joint-opt}, we solve the optimization problem at hand under multiple constraints using the underlying OOVAMP algorithm. Lastly, numerical results are presented in \mbox{Section \ref{sec:numeric-res}} before drawing out the conclusion in Section \ref{sec:conclusion}.

\vspace{0.2cm}
\noindent
\textbf{Notations}: Lowercase letters (e.g., $r$) denote scalar variables. 
Vectors are denoted by small boldface letters (e.g., $\zs$) and the $k$-th element of $\zs$ is denoted as $z_k$. Exponent on a vector (e.g., $\zs^n$) denotes component-wise exponentiation on every element of the vector. Capital boldface letters (e.g., $\A$) are used to denote matrices, while $a_{ik}$ and $\bm{a}_i$ stand, respectively, for the $(i,k)$-th entry and the $i$-th column of $\A$. The zero matrix of size $M \times N$ is denoted as $\0_{M \times N}$. $\C^{ M \times N}$ stands for the set of matrices of size $M \times N$ with complex elements and $\A^{-k}$ means $\left(\A^{-1}\right)^k$. $\tn{Rank}(\A)$ and $\tn{Tr}(\A)$, return, respectively, the rank and the trace of any matrix $\A$. We also use $\|.\|?2, \ \|.\|?F, \ (.)^*, \ (.)^\tr, \ (.)^\hr$ to denote the $\mathcal{L}?2$ norm, Frobenius norm, the conjugate, the transpose, and the conjugate transpose operators, respectively. The operator $<.>$ returns the empirical average of all the elements/entries of any vector or matrix. Moreover, $\tn{vec}(.)$ and $\tn{unvec}(.)$ denote vectorization of a matrix and unvectorization of a vector back to its original matrix form, respectively. $\tn{Diag}(.)$ operates on a vector and generates a diagonal matrix by placing that vector in the diagonal whereas $\tn{diag}(.)$ operates on a matrix and returns its main diagonal in a vector. The statistical expectation is denoted as $\E\lbrace.\rbrace$. A random vector with complex normal distribution is represented by $\xxn \sim \CG\N(\xs;\us,\rt)$, where $\us$ and $\rt$ denote its mean and covariance matrix, respectively. Similarly, a random matrix with complex normal distribution is represented by $\xn \sim \CG\M\N(\x;\m,\U, \V)$, where $\m$, $\U$ and $\V$ denote its mean and covariance matrices among its rows and columns, respectively. The imaginary unit is represented by $\js=\sqrt{-1}$ and the $\angle (.)$ operator returns the angle of any complex number. The proportional relationship between any two entities (functions or variables) is denoted by $\varpropto$. Lastly, the operators $\otimes$, $\odot$ and $\ast$ stand for the Kronecker, the Hadamard, and the column-wise Khatri-Rao products, respectively.

\section{System Model, Assumptions, and Problem Formulation}
\label{sec:system-model}
\begin{figure}
\bc
\scalebox{1.2}{
\begin{picture}(250,180)
\put(0,0){\includegraphics[scale=0.46]{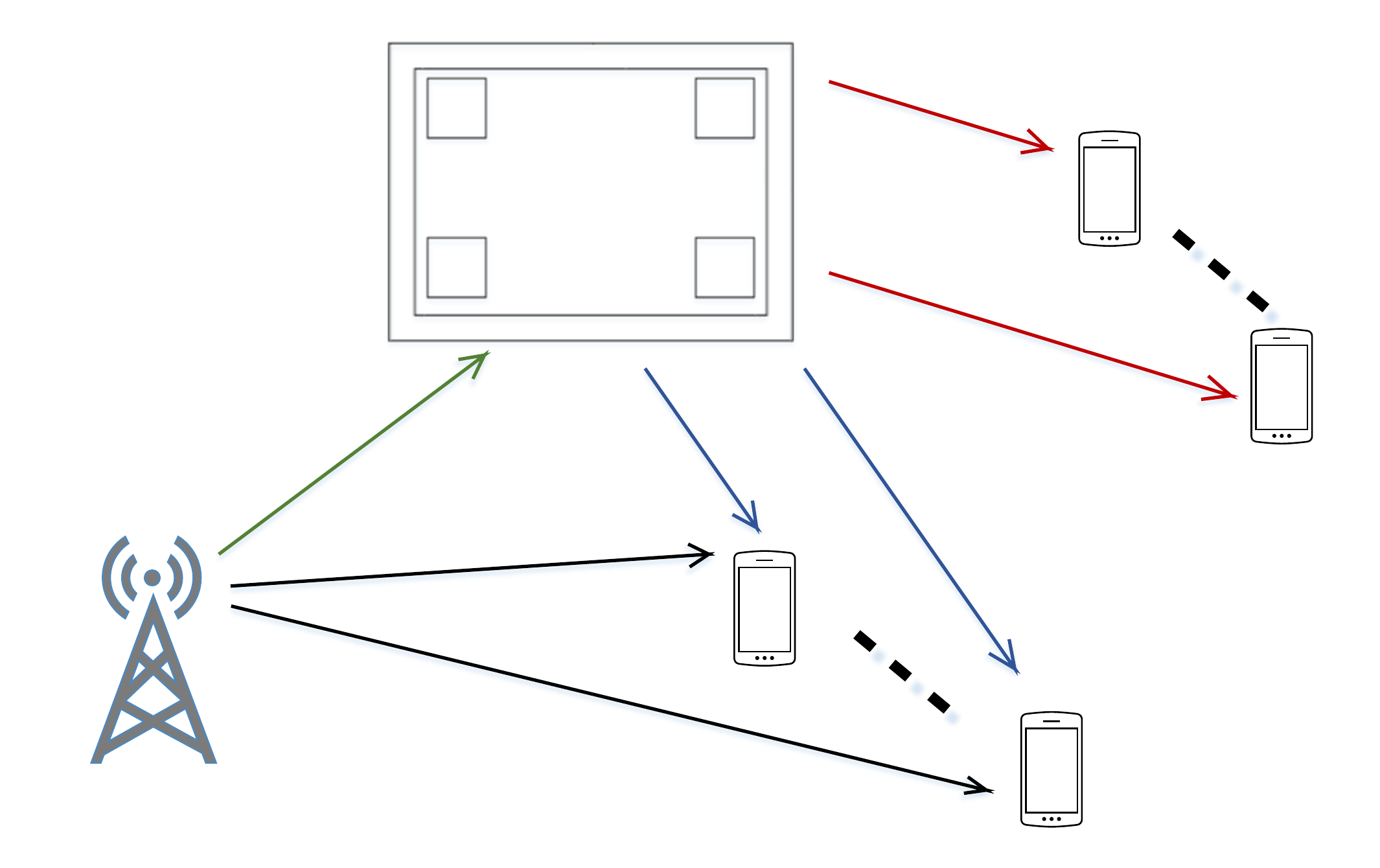}}
\put(24,11){BS}
\put(80,170){MIS $K$ elements}
\put(148,32){\footnotesize{User $1$}}
\put(202,0){\footnotesize{User $B$}}
\put(90,51){$\hs_{\tn{b-ub},1}$}
\put(90,31){$\hs_{\tn{b-ub},B}$}
\put(90,81){$\h?{b-s}$}
\put(148,81){$\hs_{\tn{s-ub},1}$}
\put(184,81){$\hs_{\tn{s-ub},B}$}
\put(90,135){{\huge $\vdots$}}
\put(146,135){{\huge $\vdots$}}
\put(109,151){{\huge $\cdots$}}
\put(109,115){{\huge $\cdots$}}
\put(109,133){{\huge $\ddots$}}
\put(214,119){\footnotesize{User $1$}}
\put(244,77){\footnotesize{User $R$}}
\put(184,142){$\hs_{\tn{s-us},1}$}
\put(184,104){$\hs_{\tn{s-us},R}$}
\end{picture}
}
\caption{MIS-assisted multi-user MIMO system in which the MIS is concurrently used for beamforming and data embedding.}
\label{fig:sys-model}
\ec
\end{figure}
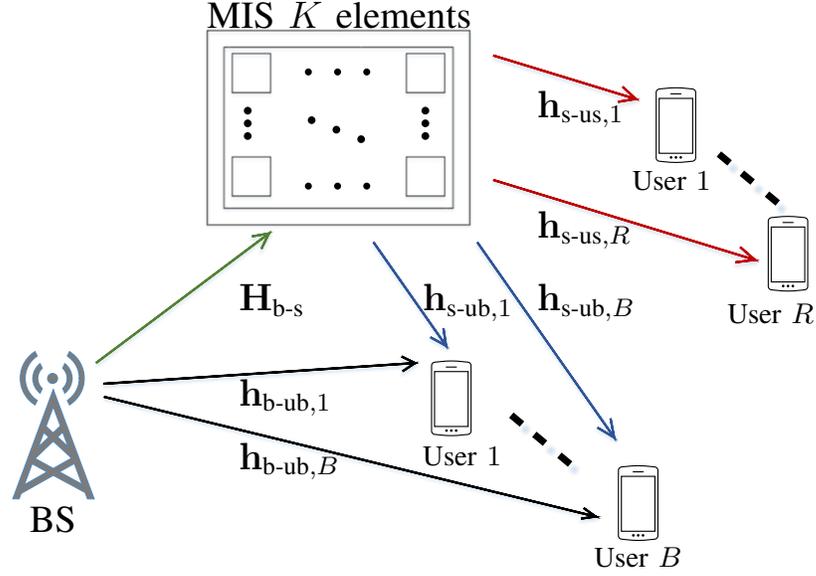

Consider a BS that is equipped with $N$ antenna elements that is serving a total of $M$ single-antenna users (in the downlink) with the help of a MIS that has $K>M$ reflective elements. Also consider a scheme in which the data for $B$ out of the total $M$ users are directly modulated/encoded by the BS (in baseband). Those $B$ users ($B < N$) are referred to as the \textit{BS-served} users. For the remaining $R = M-B$ users, the BS will simply send a known/reference  signal which will then be modulated by appropriately phase-shifting it using the MIS reflective elements. For this reason, we call those $R$ users as the \textit{MIS-served} users although, strictly speaking, both types of users are being served by the BS\footnote{Indeed, although being applied at the MIS,  the information-bearing phase shifts are designed centrally at the BS as function of the users' data.}. The goal is to optimally design the MIS phase shifts not only to modulate the data for the MIS-served users but also to passively beamform the signals for the BS-served users.
As illustrated by Fig. \ref{fig:sys-model}, each $b$-th BS-served user has a direct link to the BS which is expressed by a channel vector $\hs_{\tn{b-ub},b} \in \C^{N}$. The BS-user channel vector for each $r$-th MIS-served user\footnote{Although the MIS-served users do not receive data transmitted by the BS, they will experience interference from the direct BS-user link of the BS-served users.} is denoted by $\hs_{\tn{b-us},r}\in \C^{N}$.
The surface-user channels for the $b$-th BS-served user and the $r$-th MIS-served user are denoted, respectively, by $\hs_{\tn{s-ub},b} \in \C^{K}$ and $\hs_{\tn{s-us},r} \in \C^{K}$. 
Let $\h?{b-s} \in \C^{K \times N}$ denote the channel matrix of the MIMO MIS-BS link with $\tn{Rank}(\h?{b-s})\geq B$. The signal received at the MIS is phase-shifted by a diagonal matrix $\tn{Diag}(\bm\upsilon) \in \C^{K \times K}$, where $\bm\upsilon \in \C^K$ is the phase-shift vector that might be subject to various constraints in practice. In particular, under the unimodular constraint, we have $\abs*{\upsilon_k}=1$ for $k=1,\cdots,K$, or equivalently,  $\upsilon_k=e^{\js\theta_k}$ for some phase shift $\theta_k\in[0,2\pi]$. 
Let  $\h?{b-ub} = \left[\hs_{\tn{b-ub},1}, \hs_{\tn{b-ub},2}, \cdots, \hs_{\tn{b-ub},B}\right] \in \C^{N \times B}$, $\h?{b-us} = \left[\hs_{\tn{b-us},1}, \hs_{\tn{b-us},2}, \cdots, \hs_{\tn{b-us},R}\right] \in \C^{N \times R}$, $\h?{s-ub} = [\hs_{\tn{s-ub},1}, \hs_{\tn{s-ub},2}, \cdots, \hs_{\tn{s-ub},B}]  \in \C^{K \times B}$, and $\h?{s-us} = [\hs_{\tn{s-us},1}, \hs_{\tn{s-us},2}, \cdots, \hs_{\tn{s-us},R}] \in \C^{K \times R}$.
For mathematical convenience, we gather the channels between the BS (resp. MIS) and all users in a single matrix $\h?{b-u}$ (resp. $\h?{s-u}$):
\begin{align}
     \h?{b-u}&=\left[\h?{b-ub} ~ \h?{b-us}\right] \in \C^{N \times M},\\
    \h?{s-u}&=\left[\h?{s-ub} ~ \h?{s-us}\right]~ \in \C^{K \times M}.
\end{align}
 Let $\bar{\mathsf{y}}_m$ be the \textit{noiseless} signal received by the $m$-th user and define  $\bar{\ys} \triangleq [\bar{\mathsf{y}}_1, \bar{\mathsf{y}}_2,\cdots,\bar{\mathsf{y}}_M]^{\mathsf{T}}$. In the sequel, without loss of generality, we assume that $\{\bar{\mathsf{y}}_1,\bar{\mathsf{y}}_2,\cdots,\bar{\mathsf{y}}_B\}$ and $\{\bar{\mathsf{y}}_{B+1},\bar{\mathsf{y}}_{B+2},\cdots,\bar{\mathsf{y}}_{M}\}$ pertain to the BS-served and MIS-served users, respectively.  Then, $\bar{\ys}$ can be decomposed as $\bar{\ys}\,=\,\bar{\ys}?b\,+\,\bar{\ys}?s$ where $\bar{\ys}?b$ (resp. $\bar{\ys}?s$) is the information-bearing signal intended to the BS-served (resp. MIS-served) users which are given by:
\begin{align}
   \label{eqn:noiseless-bs-served} \bar{\ys}?b&~=~\h?{s-u}^\hr\tn{Diag}(\bm\upsilon)\h?{b-s}\f\ssn?b~+~ \h?{b-u}^\hr\f\ssn?b,\\
     \label{eqn:noiseless-mis-served}\bar{\ys}?s&~=~\h?{s-u}^\hr\tn{Diag}(\bm\upsilon)\h?{b-s}\left(\sqrt{P?s}\vs?b\right).
\end{align}
In \eqref{eqn:noiseless-bs-served}, $\ssn?b \sim \CG\N(\ssb;\0,\I_B)$ is the unknown symbol vector being transmitted by the BS (to the B-served users), and $\bm\upsilon\,=\,[\upsilon_1,\upsilon_2,\cdots,\upsilon_K]^{\mathsf{T}} \in \C^K$ is a vector that gathers all the phase shifts used by the MIS. Moreover,  $\f \in \C^{N \times B}$ is the active precoding matrix that is used for beamforming purposes at the BS, which satisfies $\norm{\f}?F^2=P?b$, where $P?b$ is the fraction of power being allocated to the BS-served users.
In (\ref{eqn:noiseless-mis-served}), $\sqrt{P?s}\vs?b$ is a separate reference (which is constant and known) being transmitted by the BS towards the MIS with $\norm{\vs?b}_2^2=1$ and $P?s$ is the fraction of transmit power being allocated to the MIS-served users.  The total transmit power is denoted by $P=P?b+P?s$. 
Now, let $\ws?b \sim \CG\N(\ws;~\0,~\sigma?w^2\I_B)$ and $\ws?s \sim \CG\N(\ws;~\0,~\sigma?w^2\I_R)$ denote the additive white Gaussian noise (AWGN) vectors pertaining to the BS-served and MIS-served users, respectively. Therefore, the \textit{noisy} received signal at all the users, $\ys$, is given by:
\begin{equation}
\label{eqn:sys-model-vec-simp}
\ys=\alpha?b\bar{\ys}?b + \alpha?s\bar{\ys}?s + \big[\alpha?b\ws?b^\tr, ~ \alpha?s\ws?s^\tr\big]^\tr,
\end{equation}
wherein $\alpha?b$ and $\alpha?s$ are some real-valued receive scaling factors\footnote{This is a common practice in the precoding optimization literature \cite{nossek2005,jeddaprecoding}.}. They are only utilized to facilitate the optimization of the other variables (i.e., the BS precoder and the MIS phase shifts) since the receivers can blindly estimate these scalars based on the received symbol sequence \cite{nossek2005,jeddaprecoding}.
\begin{remark}
The system model in \eqref{eqn:sys-model-vec-simp} is an approximation of the exact system model which is expressed as follows:
\begin{equation}
\label{eqn:sys-model-vec-exact}
\begin{bmatrix}
\alpha?b\I_B & \0_{B \times R}\\
\0_{R \times B} & \alpha?s\I_R
\end{bmatrix} 
\Big(\bar{\ys}?b + \bar{\ys}?s + \big[\ws?b^\tr, ~ \ws?s^\tr\big]^\tr\Big).
\end{equation}
The approximation will allow better tractability for the receive scaling factors, $\alpha?b$ and $\alpha?s$, and the precoding matrix, $\f$, by decoupling $\alpha?s$ from the other two variables in the joint optimization problem solved in Section \ref{sec:joint-precod-scale}. Therefore, we use the approximate  model in \eqref{eqn:sys-model-vec-simp} to optimize the variables and then use those optimized variables together with the exact system model in \eqref{eqn:sys-model-vec-exact} to compute the performance metrics such as sum-rate in Section \ref{sec:numeric-res}.
\end{remark}
\noindent Consequently, by using \eqref{eqn:noiseless-bs-served} and \eqref{eqn:noiseless-mis-served} in \eqref{eqn:sys-model-vec-simp}, it follows that: 
\begin{equation}
\label{eqn:sys-model-vec}
\ys=\alpha?b\left(\h?{s-u}^\hr\tn{Diag}(\bm\upsilon)\h?{b-s}\f\ssn?b+ \h?{b-u}^\hr\f\ssn?b \right) + \alpha?s\left(\h?{s-u}^\hr\tn{Diag}(\bm\upsilon)\h?{b-s}\sqrt{P?s}\vs?b\right) + \big[\alpha?b\ws?b^\tr, ~ \alpha?s\ws?s^\tr\big]^\tr.
\end{equation}
Contrarily to the reference signal $\vs?b$, the optimized MIS phase-shift vector varies in space (with changes in the channel) and in time (with every transmit symbol vector). We now extend the system model in \eqref{eqn:sys-model-vec} for a transmit block of length, $L$,  as follows:
\begin{align}
\begin{split}
        \yn&=\alpha?b\Big[\h?{s-u}^\hr\tn{Diag}(\bm\upsilon_1)\h?{b-s}\f\ssn_{\tn{b},1}+ \h?{b-u}^\hr\f\ssn_{\tn{b},1},\cdots, \h?{s-u}^\hr\tn{Diag}(\bm\upsilon_L)\h?{b-s}\f\ssn_{\tn{b},L}+ \h?{b-u}^\hr\f\ssn_{\tn{b},L}\Big]\\
        &~~~~~~~~~~+\alpha?s\sqrt{P?s}\Big[\h?{s-u}^\hr\tn{Diag}(\bm\upsilon_1)\h?{b-s}\vs?b, \cdots, \h?{s-u}^\hr\tn{Diag}(\bm\upsilon_L)\h?{b-s}\vs?b\Big]\,+\,\Big[\alpha?b\wn?b^\tr, ~ \alpha?s\wn?s^\tr\Big]^\tr.
\end{split}
\end{align}
 We further gather the information symbols for the BS-served (resp. MIS-served) users in one matrix denoted as $\sn?b\triangleq [\ssn_{\tn{b},1},\cdots,\ssn_{\tn{b},L}]\in \C^{B \times L}$ (resp. $\s?s \in \C^{R \times L}$)
 and we define $\sn\triangleq\big[\sn?b^\tr,~\s?s^\tr\big]^\tr$.  We also stack the phase-shift vectors being used  at all time indices $l=1,\cdots,L$ in one single matrix $\bm\Upsilon=[\bm\upsilon_1,~ \bm\upsilon_2,\cdots,\bm\upsilon_L] \in \C^{K \times L}$.
The goal is to maximize each user's signal-to-interference-plus-noise ratio (SINR) by minimizing the error in its received symbols under the sum MMSE criterion.
A lower bound on the spectral efficiency for user $m$ can be expressed in terms of the MMSE of its received symbol error \cite{heathmimo} as follows:
\begin{equation}
C_m\<{MMSE}=\log_2\left(\frac{1}{\tn{MMSE}_m}\right).
\end{equation}
The MSE of the received symbol for user $m$ and time index $l$ is given by $\E_{\ysn_{ml},\ssf_{ml}}\left\lbrace\abs*{\ysn_{ml}-\ssf_{ml}}^2\right\rbrace$, and for $M$ users and a transmit block of length $L$, the sum symbol MSE is given by:
\begin{equation}
\ds\sum_{m=1}^M\sum_{l=1}^L\E_{\ysn_{ml},\ssf_{ml}}\left\lbrace\abs*{\ysn_{ml}-\ssf_{ml}}^2\right\rbrace \ = \ \E_{\yn,\sn}\left\lbrace\norm{\yn-\sn}?F^2\right\rbrace.
\end{equation}
Thus, our optimization problem under the sum MMSE criterion can be formulated as follows:
\begin{subequations}
\label{eqn:mmse-opt}
\begin{align}
\label{eqn:mmse-opta}
\ds\argmin_{\alpha?s, \alpha?b, \f, {\mathbf\Upsilon}} \quad & \E_{\yn,\sn}\left\lbrace\norm{\yn-\sn}?F^2\right\rbrace,\\
\label{eqn:mmse-optb}
\tn{subject to} \quad & \norm{\f}?F^2=P?b,\\
& |\upsilon_{kl}|=1, \quad k=1,2, \cdots, K, ~~ l=1,2,\cdots,L.
\end{align}
\end{subequations}
We take the expectation involved in \eqref{eqn:mmse-opta} with respect to (w.r.t.) the random matrices $\sn$, $\wn?b$, and $\wn?s$ to further simplify the objective function (see \textbf{Appendix \ref{apd:A}}) thereby resulting in the following optimization problem:

\begin{subequations}
\label{eqn:exp-mmse-opt}
\begin{align}
\label{eqn:exp-mmse-opta}
\begin{split}
    \ds\argmin_{\alpha?s, \alpha?b, \f, {\mathbf\Upsilon}} \quad &
    \bigg\|\alpha?b\Big[\h?{s-u}^\hr\tn{Diag}(\bm\upsilon_1)\h?{b-s}\f+ \h?{b-u}^\hr\f,\cdots, \h?{s-u}^\hr\tn{Diag}(\bm\upsilon_L)\h?{b-s}\f+ \h?{b-u}^\hr\f\Big]\\
    &-\Big[[\I_{B,1}, \ \0_{B\times R,1}]^\tr, \cdots, [\I_{B,L}, \ \0_{B\times R,L}]^\tr \Big]\bigg\|?F^2\\
    &+\bigg\|\alpha?s\sqrt{P?s}\Big[\h?{s-u}^\hr\tn{Diag}(\bm\upsilon_1)\h?{b-s}\vs?b, \cdots, \h?{s-u}^\hr\tn{Diag}(\bm\upsilon_L)\h?{b-s}\vs?b\Big]-\left[\0_{L \times B} \ \s?s^\tr\right]^\tr\bigg\|?F^2\\
    &+LB\sigma?w^2\alpha?b^2 + LR\sigma?w^2\alpha?s^2.
\end{split}
\\
\label{eqn:exp-mmse-optb}
\tn{s.t.} \quad & \|\f\|?F^2=P?b,\\
\label{eqn:exp-mmse-optc}
& |\upsilon_{kl}|=1, \quad k=1,2, \cdots, K, ~~ l=1,2,\cdots, L.
\end{align}
\end{subequations}
We shall denote the objective function in \eqref{eqn:exp-mmse-opta} by $f(\alpha?s, \alpha?b, \f, {\mathbf\Upsilon})$ throughout the rest of the paper.
The optimization problem in \eqref{eqn:exp-mmse-opt} is non-convex due to the unimodular constraint on the MIS phase-shifts in \eqref{eqn:exp-mmse-optc}. We aim to solve the problem by utilizing the OOVAMP algorithm in the same way as done in \cite{haseebirs}.

The OOVAMP~\cite{haseebirs} is a low-complexity algorithm which is designed to solve optimization problems involving a linear objective function and non-linear constraints \cite{ranganvamp,haseebirs}.
The modular structure of OOVAMP decouples the constraint from the objective function. This favorable property allows OOVAMP to optimize the same objective function under different constraints by modifying the OOVAMP module that satisfies the constraint (simple scalar functions). OOVAMP has better convergence properties than other iterative algorithms (e.g., alternating direction method of multipliers [ADMM]) \cite{manvamp} because of its automatic calculation of the optimal stepsize in every iteration.

\section{Optimization Oriented VAMP for Matrices}
\label{sec:vamp-opt}
OOVAMP is a recently developed extension \cite{haseebirs} of the standard max-sum VAMP algorithm\cite{ranganvamp} which solves constrained optimization problems involving linear objective functions under both linear and non-linear constraints.  Moreover, asymptotic optimality can be claimed for the computed solution under certain mild conditions using state evolution arguments. In this section, we present an extended version of the OOVAMP algorithm to optimize matrices involving linear mixing. Given the knowledge of two matrices, $\A \in \C^{M \times N}$ and $\z \in \C^{M \times K}$, the OOVAMP algorithm solves the following optimization problem:
\begin{subequations}
\label{eqn:vamp-opt}
\begin{align}
\label{eqn:vamp-opta}
\ds\argmin_{\x~\in~\C^{N \times K}} \quad & \norm{\A\x-\z}?F^2\\
\label{eqnd:vamp-optb}
\tn{s.t.} \quad & f_{ik}(x_{ik}) =0 \quad i=1, \cdots, N, \quad k=1, \cdots, K.
\end{align}
\end{subequations}
The algorithm consists of the following two modules.
\subsubsection{Linear MAP Estimator}
At iteration $t$, the linear MAP (LMAP) estimator receives extrinsic information (message) from the separable (i.e., entry-wise)  MAP projector of $\x$ in the form of a mean matrix, $\rt_{t-1}$, and a common scalar precision, $\gamma_{t-1}$. Then, under the Gaussian prior, $\CG\M\N\Big(\x;\rt_{t-1},\gamma_{t-1}^{-1}\I_N,\I_K\Big)$, it computes the LMAP estimate, $\bar{\x}_t$, along with the associated posterior precision, $\bar{\gamma}_t$, as follows:
\begin{align}
\label{eqn:lmmse-est}
\bar{\x}_t&~=~\left(\A^\hr\A+\gamma_{t-1}\I_N\right)^{-1}\left(\A^\hr\z+\gamma_{t-1}\rt_{t-1}\right), \\
\label{eqn:lmmse-var}
\bar{\gamma}_t&~=~N\tn{Tr}\left(\left[\A^\hr\A+\gamma_{t-1}\I_N\right]^{-1}\right)^{-1}.
\end{align}
The extrinsic information on $\x$ is updated as:
$$\CG\M\N(\x;\bar{\x}_t,\bar{\gamma}_t^{-1}\I_N,\I_K)/\CG\M\N(\x;\rt_{t-1},\gamma_{t-1}^{-1}\I_N,\I_K),$$
and then sent back in the form of a mean matrix, $\widetilde{\rt}_t=\left(\bar{\x}_t\bar{\gamma}_t-\rt_{t-1}\gamma_{t-1}\right)/\left(\bar{\gamma}_t-\gamma_{t-1}\right)$, and a scalar precision, $\widetilde{\gamma}_t=\bar{\gamma}_t-\gamma_{t-1}$, to the separable MAP projector of $\x$.
\subsubsection{Separable MAP Projector}
Because the constraint on $\x$ is component-wise, the constraint on its entries, $x_{ik}$, is modeled as a prior with some precision, $\gamma?p$, i.e., $p_{\xsn}(x_{ik}) \varpropto \exp \left(-\gamma?p\abs*{f_{ik}(x_{ik})}^2\right)$ with $\gamma?p \to \infty$, which results in the following prior distribution on $\x$:
\begin{equation}
p_{\xn}(\x)=\ds\prod_{i=1}^N\prod_{k=1}^Q p_{\xsn}(x_{ik}).
\end{equation}
This module computes the MAP estimate, $\widehat{\x}_t$, of $\x$ from the joint distribution\\ $p_{\xn}(\x)\CG\M\N\Big(\x;\widetilde{\rt}_t,\widetilde{\gamma}^{-1}_t\I_N,\I_K\Big)$. The MAP estimate can be computed through a component-wise projector function as follows:
\begin{equation}
\widehat{x}_{ik,t}~=~g_{ik}(\widetilde{r}_{ik,t},\widetilde{\gamma}_t)~\triangleq~ \ds\argmin_{x_{ik}}\left[\widetilde{\gamma}_t\abs*{x_{ik}-\widetilde{r}_{ik,t}}^2-\ln p_{\xsn}(x_{ik})\right],
\end{equation}
or equivalently:
\begin{equation}
\label{eqn:gen-projector}
g_{ik}(\widetilde{r}_{ik,t},\widetilde{\gamma}_t)= \argmin_{x_{ik}}\left[\widetilde{\gamma}_t\abs*{x_{ik}-\widetilde{r}_{ik,t}}^2+\gamma?p\abs*{f_{ik}(x_{ik})}^2\right].
\end{equation}
The parameter $\gamma?p$ in \eqref{eqn:gen-projector} accounts for the weight given to the prior on $x_i$ inside the scalar MAP optimization. Therefore, taking $\gamma_p \to \infty$ enforces the constraint.
The derivative of the scalar MAP projector w.r.t. $\widetilde{r}_{ik,t}$ is given by \cite{ranganvamp}:
\begin{equation}
\label{eqn:denoiser-der}
g_{ik}'(\widetilde{r}_{ik,t},\widetilde{\gamma}_t)~\triangleq~\frac{\partial g_{ik}(\widetilde{r}_{ik,t},\widetilde{\gamma}_t)}{\partial\widetilde{r}_{ik,t}}
~=~\frac{1}{2}\left(\frac{\partial g_{ik}\left(\widetilde{r}_{ik,t},\widetilde{\gamma}_t\right)}{\partial\Re\left\lbrace\widetilde{r}_{ik,t}\right\rbrace}-\js\frac{\partial g_{ik}\left(\widetilde{r}_{ik,t},\widetilde{\gamma}_t\right)}{\partial\Im\left\lbrace\widetilde{r}_{ik,t}\right\rbrace}\right)
~=~\widetilde{\gamma}_t\widehat{\gamma}_t,
\end{equation}
where $\widehat{\gamma}_t$ is the posterior precision. The matrix-valued projector function and its derivative are defined as follows:
\begin{equation}
\label{eqn:matrix-pro}
\g(\widetilde{\rt}_t,\widetilde{\gamma}_t)~\triangleq~
\begin{bmatrix}
g_{11}(\widetilde{r}_{11,t},\widetilde{\gamma}_t) & g_{12}(\widetilde{r}_{12,t},\widetilde{\gamma}_t) & \cdots &
g_{1K}(\widetilde{r}_{1K,t},\widetilde{\gamma}_t)\\
g_{21}(\widetilde{r}_{21,t},\widetilde{\gamma}_t) & g_{22}(\widetilde{r}_{22,t},\widetilde{\gamma}_t) & \cdots &
g_{2K}(\widetilde{r}_{2K,t},\widetilde{\gamma}_t)\\
\vdots & \vdots & \ddots & \vdots\\
g_{N1}(\widetilde{r}_{N1,t},\widetilde{\gamma}_t) & g_{N2}(\widetilde{r}_{N2,t},\widetilde{\gamma}_t) & \cdots &
g_{NK}(\widetilde{r}_{NK,t},\widetilde{\gamma}_t)
\end{bmatrix},
\end{equation}

\begin{equation}
\label{eqn:matrix-pro-der}
\g'(\widetilde{\rt}_t,\widetilde{\gamma}_t)~\triangleq~
\begin{bmatrix}
g'_{11}(\widetilde{r}_{11,t},\widetilde{\gamma}_t) & g'_{12}(\widetilde{r}_{12,t},\widetilde{\gamma}_t) & \cdots &
g'_{1K}(\widetilde{r}_{1K,t},\widetilde{\gamma}_t)\\
g'_{21}(\widetilde{r}_{21,t},\widetilde{\gamma}_t) & g'_{22}(\widetilde{r}_{22,t},\widetilde{\gamma}_t) & \cdots &
g'_{2K}(\widetilde{r}_{2K,t},\widetilde{\gamma}_t)\\
\vdots & \vdots & \ddots & \vdots\\
g'_{N1}(\widetilde{r}_{N1,t},\widetilde{\gamma}_t) & g'_{N2}(\widetilde{r}_{N2,t},\widetilde{\gamma}_t) & \cdots &
g'_{NK}(\widetilde{r}_{NK,t},\widetilde{\gamma}_t)
\end{bmatrix}.
\end{equation}
Similar to the LMAP module, the MAP projector module computes an extrinsic mean matrix, $\rt_t=\left(\widehat{\x}_t\widehat{\gamma}_t-\widetilde{\rt}_t\widetilde{\gamma}_t\right)/\left(\widehat{\gamma}_t-\widetilde{\gamma}_t\right)$, and a scalar precision, $\gamma_t=\widehat{\gamma}_t-\widetilde{\gamma}_t$, and sends them back to the LMAP module for the next iteration. The process is repeated until convergence.

It is worth mentioning that the extrinsic parameters, i.e., the extrinsic mean matrix and the scalar precision, calculated by each module act as a Gaussian prior on the subsequent estimate of the adjacent module, thus making OOVAMP parameter-free. Another major advantage of OOVAMP is that it decouples the constraint from the objective function and also allows the projector function to be separable. While the LMAP module optimizes the objective function with no constraints, the latter are enforced by the projector function. This modular property makes OOVAMP an attractive algorithm for solving optimization problems involving linear mixing and under various component-wise constraints. The algorithmic steps of OOVAMP are shown in \textbf{Algorithm \ref{algo:opt-vamp}}.
\begin{algorithm}
\caption{Optimization-oriented max-sum matrix VAMP}
\label{algo:opt-vamp}
\mbox{\small Given $\A\in\C^{M \times N}$, $\z \in \C^{M \times Q}$, a precision tolerance $(\epsilon)$ and a maximum number of iterations $(T?{MAX})$}
\vspace{-20pt}
\begin{algorithmic}[1]
\STATE Initialize $\rt_0$, $\gamma?{0}\geq0$ and $t\leftarrow 1$
\REPEAT
\STATE \vspace{2pt}// LMAP.
\STATE $\bar{\x}_t=\left(\A^\hr\A+\gamma_{t-1}\I_N\right)^{-1}\left(\A^\hr\z+\gamma_{t-1}\rt_{t-1}\right)$
\STATE $\bar{\gamma}_t=N\tn{Tr}\left(\left[\A^\hr\A+\gamma_{t-1}\I_N\right]^{-1}\right)^{-1}$
\STATE $\widetilde{\gamma}_t=\bar{\gamma}_t-\gamma_{t-1}$
\STATE $\widetilde{\rt}_t=\widetilde{\gamma}_t^{-1}\left(\bar{\x}_t\bar{\gamma}_t-\rt_{t-1}\gamma_{t-1}\right)$ \vspace{5pt}

\STATE // Separable MAP Projector
\STATE $\widehat{\x}_t=\g\Big(\widetilde{\rt}_t,\widetilde{\gamma}_t\Big)$
\STATE $\widehat{\gamma}_t=\widetilde{\gamma}_t^{-1}\left\langle \g'\Big(\widetilde{\rt}_t,\widetilde{\gamma}_t\Big)\right\rangle$
\STATE $\gamma_t=\widehat{\gamma}_t-\widetilde{\gamma}_t$
\STATE $\rt_t=\gamma_t^{-1}\Big(\widehat{\gamma}_t\widehat{\x}_t-\widetilde{\gamma}_t\widetilde{\rt}_t\Big)$ 
\STATE $t\leftarrow t+1$
\vspace{2pt}
\UNTIL $\norm{\widehat{\x}_t-\widehat{\x}_{t-1}}_2^2\leq \epsilon\norm{\widehat{\x}_{t-1}}_2^2$ or $t>T?{MAX}$
\RETURN $\widehat{\x}_t$
\end{algorithmic}
\end{algorithm}

\section{OOVAMP-Based Solution for the Optimization Problem}
\label{sec:joint-opt}

In this section, we apply the OOVAMP algorithm, introduced in Section \ref{sec:vamp-opt}, to simultaneously optimize the matrix of phase shifters, ${\mathbf\Upsilon}$, the optimal precoding matrix $\f$, and the scaling factors, $\alpha?b$ and $\alpha?s$ (i.e., to solve the optimization problem in  (\ref{eqn:exp-mmse-opt})). We follow the optimization procedure presented in \cite{haseebirs}, and decouple the joint optimization problem into two sub-problems through alternating optimization. In one side, we optimize ${\mathbf\Upsilon}$ by utilizing OOVAMP and, on the other side, we find the optimal transmit precoding, $\f$, and scalars $\alpha?b$ and $\alpha?s$.
It is a simple iterative approach that optimizes a subset of all variables at a time while fixing the other set of variables and the process is repeated until convergence. More specifically, we divide the optimization problem in \eqref{eqn:mmse-opt} into the following two sub-optimization problems:

\begin{enumerate}
\item 
\begin{subequations}
\label{eqn:ao-phase-mat}
\begin{align}
\label{eqn:ao-phase-mata}
\widehat{\bm\Upsilon}=\ds\argmin_{{\mathbf\Upsilon}} \quad & 
f(\alpha?s, \alpha?b, \f, {\mathbf\Upsilon})\\
\label{eqn:ao-phase-matb}
\tn{s.t.} \quad & |\upsilon_{kl}|=1, ~~ k=1,2, \cdots, K, ~~ l=1,2, \cdots, L.
\end{align}
\end{subequations}
\item
\begin{subequations}
\label{eqn:ao-precod-mat}
\begin{align}
\label{eqn:ao-precod-mata}
\argmin_{\alpha?s,\alpha?b, \f} \quad & 
f(\alpha?s, \alpha?b, \f, \widehat{\mathbf\Upsilon})\\
\label{eqn:ao-precod-matb}
\tn{s.t.} \quad & \norm{\f}?F^2 =P?b.
\mspace{210mu}\end{align}
\end{subequations}
\end{enumerate}

\subsection{Optimizing the MIS Phase Shifts}
\label{sec:sub-opt-phase}
Here, we derive the OOVAMP modules (i.e., LMAP estimator and separable projector function) to solve the sub-optimization problem in \eqref{eqn:ao-phase-mat} which is restated explicitly as follows:
\begin{subequations}
\label{eqn:opt-phase-mat}
\begin{align}
\label{eqn:opt-phase-mata}
\begin{split}
    \ds\argmin_{{\mathbf\Upsilon}} \quad &
    \bigg\|\alpha?b\Big[\h?{s-u}^\hr\tn{Diag}(\bm\upsilon_1)\h?{b-s}\f+ \h?{b-u}^\hr\f,\cdots, \h?{s-u}^\hr\tn{Diag}(\bm\upsilon_L)\h?{b-s}\f+ \h?{b-u}^\hr\f\Big]\\
    &-\Big[[\I_{B,1}, \ \0_{B\times R,1}]^\tr, \cdots, [\I_{B,L}, \ \0_{B\times R,L}]^\tr \Big]\bigg\|?F^2\\
    &+\bigg\|\alpha?s\sqrt{P?s}\Big[\h?{s-u}^\hr\tn{Diag}(\bm\upsilon_1)\h?{b-s}\vs?b, \cdots, \h?{s-u}^\hr\tn{Diag}(\bm\upsilon_L)\h?{b-s}\vs?b\Big]-\left[\0_{L \times B} \ \s?s^\tr\right]^\tr\bigg\|?F^2\\
    &+LB\sigma?w^2\alpha?b^2 + LR\sigma?w^2\alpha?s^2
\end{split}
\\
\label{eqn:opt-phase-matb}
\tn{s.t.} \quad & |\upsilon_{kl}|=1, ~~ k=1,2, \cdots, K, ~~ l=1,2, \cdots, L.
\end{align}
\end{subequations}
Next, we re-express the objective function in \eqref{eqn:opt-phase-mata} in a form that is similar to the general OOVAMP objective function in \eqref{eqn:vamp-opta}. In fact, by introducing the following matrices:
\begin{align}
\label{eqn:opt-lmap-mat-a}
\A &=\alpha?b\h?{s-u}^\hr,&(M \times K),\\
\B &=(\h?{b-s}\f)^\tr,&(B \times K),\\
\D &=\left[\bs_1 \otimes \as_1, \cdots, \bs_K \otimes \as_K\right], &(MB \times K),\\
\m &=\sqrt{P?s}\alpha?s\h?{s-u}^\hr\tn{Diag}(\h?{b-s}\vs?b),&(M \times K),\\
\x &=\big[\tn{vec}\left([\I_{B}, \ \0_{B\times R}]^\tr-\alpha?b\h?{b-u}^\hr\f\right), \cdots, \tn{vec}\left([\I_{B}, \ \0_{B\times R}]^\tr-\alpha?b\h?{b-u}^\hr\f\right)\big],\hspace{-4pt}&(MB \times L),\\
\label{eqn:opt-lmap-mat-z}
\z &=\left[\0_{L \times B}, \ \s?s^\tr\right]^\tr, &(M \times L),
\end{align}
we show in \textbf{Appendix \ref{apd:B}} that the optimization problem in \eqref{eqn:opt-phase-mat} can be rewritten as follows:
\begin{subequations}
\label{eqn:opt-phase-mat-short}
\begin{align}
\label{eqn:opt-phase-mata-short}
\ds\argmin_{{\mathbf\Upsilon}} \quad & \norm{\D{\bm\Upsilon}-\x}?F^2 ~+~ \norm{\m{\bm\Upsilon}-\z}?F^2~+~LB\sigma?w^2\alpha?b^2~+~ LR\sigma?w^2\alpha?s^2\\
\label{eqn:opt-phase-matb-short}
\tn{s.t.} \quad & |\upsilon_{kl}|=1, ~~ k=1,2, \cdots, K, ~~ l=1,2, \cdots, L.
\end{align}
\end{subequations}
The steps to derive both OOVAMP modules are detailed in the sequel.

\subsubsection{LMAP Estimator} The LMAP module performs the minimization of the objection function in \eqref{eqn:opt-phase-mata-short} under the Gaussian prior, $\CG\M\N\Big({\bm\Upsilon};\rt_{t-1},\gamma_{t-1}^{-1}\I_K,\I_L\Big)$, by solving the following unconstrained optimization problem:
\begin{equation}
\label{eqn:opt-map-opt}
\ds\argmin_{{\bm\Upsilon}}\quad \frac{1}{2}\norm{\D{\bm\Upsilon}-\x}?F^2 + \frac{1}{2}\norm{\m{\bm\Upsilon}-\z}?F^2 + \frac{\gamma_{t-1}}{2}\norm{{\bm\Upsilon}-\rt_{t-1}}?F^2.
\end{equation}
The solution (i.e., the LMAP estimate and the associated posterior precision) to the optimization problem in \eqref{eqn:opt-map-opt} is given as follows:
\begin{align}
\label{eqn:opt-lmap-est}
\bar{\bm\Upsilon}_t&~=~\left(\D^\hr\D + \m^\hr\m+\gamma_{t-1}\I_K\right)^{-1} \left(\D^\hr\x +\m^\hr\z+\gamma_{t-1}\rt_{t-1}\right),\\
\label{eqn:opt-lmap-var}
\bar{\gamma}_t&~=~K\tn{Tr}\left(\left[\D^\hr\D + \m^\hr\m+\gamma_{t-1}\I_K\right]^{-1}\right)^{-1}.
\end{align}
\subsubsection{Scalar MAP Projector} In this section, we introduce two projector functions defined in \cite{haseebirs} to satisfy the two types of constraint on the MIS reflection coefficients, i.e., $i)$ the unimodular constraint, and $ii)$ a practical constraint on the MIS phase shifts in which each antenna element is terminated by a variable reactive load. We refer the reader to \cite{haseebirs} for more details on the derivation and differences in the two projector functions.
The projector function and its derivative for the unimodular constraint is given as follows:
\begin{align}
    g_{1,kl}\left(\widetilde{r}_{kl}\right)&~=~
  \widetilde{r}_{kl}\abs*{\widetilde{r}_{kl}}^{-1},\\
  g'_{1,kl}\left(\widetilde{r}_{kl}\right)&~=~
  \frac{1}{2}\abs*{\widetilde{r}_{kl}}^{-1},
\end{align}
where the derivative is taken according to \eqref{eqn:denoiser-der}.

To optimize the MIS phase shifts under a practical constraint, we consider a reflective element that is combined with a tunable reactive load\footnote{This can be implemented for instance by an antenna array composed of omni-directional dipole elements loaded with the reactive elements in the absence of a ground plane to allow for bidirectional beamforming and not just hemispherical coverage.} instead of an ideal phase-shifter\footnote{The value $1$ is the normalized resistive part of the element impedance whereas $\chi_{kl}$ is the normalized reactive part of the antenna plus reactive termination. Accordingly $\upsilon_{kl}$ represents the induced current flowing across the antenna. We assume the antenna elements to be uncoupled which holds approximately for half-wavelength element spacing.}, i.e., \mbox{$\upsilon_{kl}=-(1+\js\chi_{kl})^{-1}$,}  where $\chi_{kl} \in \R$ is a scalar reactance value that must be optimized for each reflection coefficient. 
Under the unimodular constraint, the idealistic MIS has a full field of view (FOV) and the reflection coefficients correspond to ideal phase-shifters and are of the form $\upsilon_{kl}=e^{\js\theta_{kl}}$, where\footnote{From a practical standpoint, this assumption is difficult to realize. With the reactive-loading constraint, the assumption of a MIS with full field of view (FOV) becomes more acceptable.} $\theta_{kl} \in [0, 2\pi]$, whereas under the practical constraint we have a restriction on the possible values of the MIS phase-shifts, i.e., $\angle -(1+\js\chi)^{-1} \in \big[\tn{-}\frac{\pi}{2}, \frac{\pi}{2}\big]$. Moreover, the magnitude of each phase-shift under this constraint is always less than $1$ for any $\chi\neq 0$. Practically, this introduces the phase-dependent amplitude attenuation in the incident wave. The projector function under this new constraint is defined as:
\begin{equation}
\label{eqn:reactive-projector}
g_{2,kl}\left(\widetilde{r}_{kl},\widetilde{\gamma}\right)~\triangleq~
\ds\argmin_{\upsilon_{kl}}\left[\widetilde{\gamma}\abs*{\upsilon_{kl}-\widetilde{r}_{kl}}^2+\gamma?p\abs*{\upsilon_{kl}+\frac{1}{1+\js\chi_{kl}\<{opt}}}^2\right],
\end{equation}
where
\begin{equation}
\label{eqn:reactance-opt}
\chi_{kl}\<{opt}~=~g_3\left(\widetilde{r}_{kl}\right)~\triangleq~\ds\argmin_{\chi_{kl}} \abs*{\widetilde{r}_{kl}+\frac{1}{1+\js\chi_{kl}}}^2.
\end{equation}
The solution to the optimization problem in \eqref{eqn:reactance-opt}, the projector function, and its derivative are given by:
\begin{align}
    \label{eqn:reactance-opt-sol}
g_3\left(\widetilde{r}_{kl}\right)&~=~
\frac{1}{2\Im\left\lbrace\widetilde{r}_{kl}\right\rbrace}\left(1+2\Re\left\lbrace\widetilde{r}_{kl}\right\rbrace+\sqrt{\left(1+2\Re\left\lbrace\widetilde{r}_{kl}\right\rbrace\right)^2+4\Im\left\lbrace\widetilde{r}_{kl}\right\rbrace^2}\right),\\
g_{2,kl}\left(\widetilde{r}_{kl}\right)&~=~
  -\left(1+\js g_3\left(\widetilde{r}_{kl}\right)\right)^{-1},\\
g'_{2,kl}\left(\widetilde{r}_{kl}\right)&~=~
  \abs*{\js g'_3\left(\widetilde{r}_{kl}\right)\left(1+\js g_3\left(\widetilde{r}_{kl}\right)\right)^{-2}}.
\end{align}
The matrix valued projector functions, $\g_1(\widetilde{\rt}_t,\widetilde{\gamma}_t)$ and  $\g_2(\widetilde{\rt}_t,\widetilde{\gamma}_t)$, and their derivatives, $\g'_1(\widetilde{\rt}_t,\widetilde{\gamma}_t)$ and $\g'_2(\widetilde{\rt}_t,\widetilde{\gamma}_t)$, are obtained according to \eqref{eqn:matrix-pro} and \eqref{eqn:matrix-pro-der}.
Lastly, the constant transmitted vector by the BS, $\vs?b$, is set to the right singular vector of the matrix $\h?{b-s}$ that corresponds to the largest eigenvalue.
\begin{figure}
\bc
\scalebox{1.1}{
\begin{picture}(500,150)(-7.5,0)
\put(0,0){\includegraphics[scale=0.5]{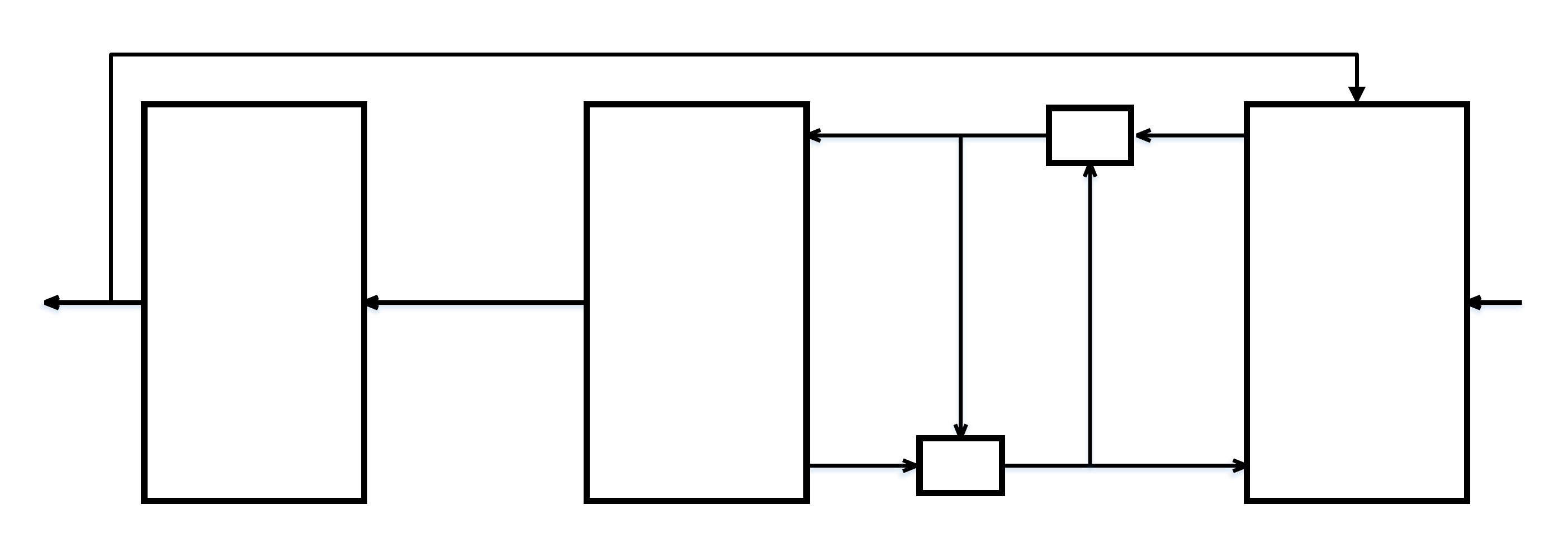}}
\put(165,70){{\footnotesize Separable}}
\put(171,58){{\footnotesize MAP}}
\put(165,47){{\footnotesize Projector}}
\put(340,64){{\footnotesize LMAP}}
\put(335,53){{\footnotesize Estimator}}
\put(276,103){{\footnotesize ext}}
\put(244,18){{\footnotesize ext}}
\put(130,65){{\footnotesize $\widehat{\bm\Upsilon}$}}
\put(382,65){{\footnotesize $\x$}}
\put(382,52){{\footnotesize $\z$}}
\put(240,108){{\footnotesize $\widetilde{\rt},\widetilde{\gamma}$}}
\put(310,108){{\footnotesize $\bar{\gamma}$}}
\put(310,94){{\footnotesize $\bar{{\bm\Upsilon}}$}}
\put(274,10){{\footnotesize $\rt,\gamma$}}
\put(216,8){{\footnotesize $\widehat{{\bm\Upsilon}}$}}
\put(216,23){{\footnotesize $\widehat{\gamma}$}}
\put(50,70){{\footnotesize Precoder}}
\put(49,58){{\footnotesize $\&$ Scalar}}
\put(48,47){{\footnotesize Optimizer}}
\put(19,65){{\footnotesize $\widehat{\f}$}}
\put(13,52){{\footnotesize $\widehat{\alpha}?b$, }}
\put(25,52){{\footnotesize $\widehat{\alpha}?s$}}
\end{picture}}
\ec
\caption{Block diagram of the proposed algorithm. The calculation of extrinsic information is performed by the ``ext" blocks.}
\label{fig:overall}
\end{figure}
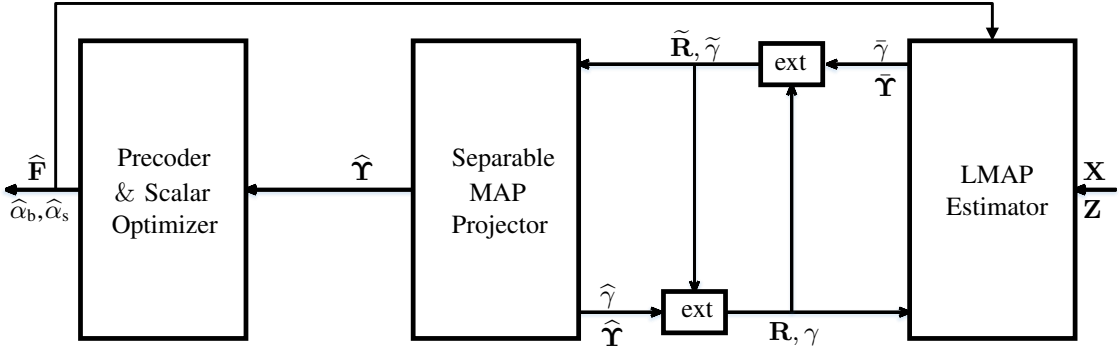

\subsection{Optimal Precoding and Scaling Factors}
\label{sec:joint-precod-scale}
The receive scaling factor, $\alpha?s$, is decoupled from the other optimization variables, i.e., $\f$ and $\alpha?b$, in the objective function \eqref{eqn:ao-precod-mata}. Therefore, it can be optimized independently of the other two variables as follows:
\begin{align}
\label{eqn:opt-alpha-s}
    \begin{split}
        \ds\argmin_{\alpha?s} \quad & \bigg\|\alpha?s\sqrt{P?s}\Big[\h?{s-u}^\hr\tn{Diag}(\bm\upsilon_1)\h?{b-s}\vs?b, \cdots, \h?{s-u}^\hr\tn{Diag}(\bm\upsilon_L)\h?{b-s}\vs?b\Big]-\left[\0_{L \times B} \ \s?s^\tr\right]^\tr\bigg\|?F^2\\
        & + LR\sigma?w^2\alpha?s^2.
    \end{split}
\end{align}
By defining the matrix:
\begin{equation}
\label{eqn:precod-mat-c}
    \CC\triangleq\sqrt{P?s}\h?{s-u}^\hr\tn{Diag}(\h?{b-s}\vs?b){\bm\Upsilon},
\end{equation}
we rewrite \eqref{eqn:opt-alpha-s} as:
\begin{equation}
    \label{eqn:opt-alpha-s-short}
    \ds\argmin_{\alpha?s} \quad \norm{\alpha?s\CC-\left[\0_{L \times B} \ \s?s^\tr\right]^\tr}?F^2 + LR\sigma?w^2\alpha?s^2.
\end{equation}
From \eqref{eqn:opt-alpha-s-short}, we establish the closed-form solution to the optimization problem in \eqref{eqn:opt-alpha-s} as follows:
\begin{equation}
    \alpha?s\<{opt}=g_4(\CC)~\triangleq~\frac{\tn{Tr}\left(\CC^\hr\left[\0_{L \times B} \ \s?s^\tr\right]^\tr+\left[\0_{L \times B} \ \s?s^\hr\right]\CC\right)}{2\left(\norm{\CC}?F^2+LR\sigma?w^2\right)}.
\end{equation}
Now, the optimal precoding matrix, $\f$, and receive scaling factor, $\alpha?b$, are obtained as a solution to the following optimization problem:
\begin{subequations}
\label{eqn:opt-precod}
\begin{align}
    \begin{split}
        \ds\argmin_{\alpha?b, \f} \quad &
        \bigg\|\alpha?b\Big[\h?{s-u}^\hr\tn{Diag}(\bm\upsilon_1)\h?{b-s}\f+ \h?{b-u}^\hr\f,\cdots, \h?{s-u}^\hr\tn{Diag}(\bm\upsilon_L)\h?{b-s}\f+ \h?{b-u}^\hr\f\Big]\\
        &-\Big[[\I_{B,1}, \ \0_{B\times R,1}]^\tr, \cdots, [\I_{B,L}, \ \0_{B\times R,L}]^\tr \Big]\bigg\|?F^2 + LB\sigma?w^2\alpha?b^2
    \end{split}\\
\tn{s.t.} \quad & \norm{\f}?F^2 =P?b. 
\end{align}
\end{subequations}

By defining the matrices:
\begin{align}
\label{eqn:precod-mat-k}
\K&~\triangleq~\ds\sum_{l=1}^L\left(\h?{s-u}^\hr\tn{Diag}({\bm\upsilon}_l)\h?{b-s}+\h?{b-u}^\hr\right)^\hr\left(\h?{s-u}^\hr\tn{Diag}({\bm\upsilon}_l)\h?{b-s}+\h?{b-u}^\hr\right)\\
\label{eqn:precod-mat-e}
\e&~\triangleq~[\I_B \ \0_{B \times R}] \ds\sum_{l=1}^L\left(\h?{s-u}^\hr\tn{Diag}({\bm\upsilon}_l)\h?{b-s}+\h?{b-u}^\hr\right),
\end{align}
the optimization problem in \eqref{eqn:opt-precod} becomes a constrained MMSE transmit precoding optimization for MIMO systems. The problem can be solved jointly by Lagrange optimization. We construct the Lagrangian function for the optimization problem in \eqref{eqn:opt-precod} as follows:
\begin{equation}
\LA(\f,\alpha?b,\lambda)=\tn{Tr}\left(\alpha?b^2\K\f\f^\hr -\alpha?b\e\f-\alpha?b\f^\hr\e^\hr\right)+LB+LB\alpha?b^2\sigma?w^2 + \lambda\left(\tn{Tr}\left(\f\f^\hr\right)-P?b\right),
\end{equation}
with $\lambda \in \R $ being the Lagrange multiplier.
The closed-form solutions for optimal $\alpha?b$ and $\f$ are given below and we refer the reader to \cite{nossek2005} for more details: 
\begin{align}
    \alpha?b\<{opt}&~=~g_5\left(\K,~\e\right)~\triangleq~\sqrt{\frac{1}{P?b}}\sqrt{\tn{Tr}\left(\left[\K+\frac{LB\sigma?w^2\I_N}{P?b}\right]^{-2}\e^\hr\e\right)},\\
    \f\<{opt}&~=~g_6\left(\K,~\e\right)~\triangleq~\frac{\sqrt{P?b}\left[\K+\frac{LB\sigma?w^2\I_N}{P?b}\right]^{-1}\e^\hr}{\sqrt{\tn{Tr}\left(\left[\K+\frac{LB\sigma?w^2\I_N}{P?b}\right]^{-2}\e^\hr\e\right)}}.
\end{align}

\begin{algorithm*}
\caption{OOVAMP-based joint optimization algorithm}
\label{algo:overall}
\mbox{\small Given $\h?{s-u}$, $\h?{b-u}$, $\h?{b-s}$, $\s?s$, a precision tolerance $(\epsilon)$, and a maximum number of iterations $(T?{MAX})$}
\vspace{-20pt}
\begin{algorithmic}[1]
\STATE  Initialize $\widehat{{\bm\Upsilon}}?0, \ \rt_0$, $\gamma_0\geq0$ and $t\leftarrow 1$, and obtain $\vs?b$ from $\h?{b-s}$
\STATE Compute $\widehat{\CC}_0,~\widehat{\K}_0$ and $\widehat{\e}_0$ by substituting $\widehat{{\bm\Upsilon}}?0$ into \eqref{eqn:precod-mat-c}, \eqref{eqn:precod-mat-k} and \eqref{eqn:precod-mat-e}.
\STATE $\widehat{\alpha}_{\tn{s},0}=g_4\left(\widehat{\CC}_0\right)$
\STATE $\widehat{\alpha}_{\tn{b},0}=g_5\left(\widehat{\K}_0,~\widehat{\e}_0\right)$
\STATE $\widehat{\f}?0=g_6\left(\widehat{\K}_0,~\widehat{\e}_0\right)$ \vspace{5pt}
\REPEAT
\STATE // LMAP Estimator
\STATE Compute matrices $\A,~\B,~\D,~\m,~\x$ and $\z$ by substituting $\widehat{\alpha}_{\tn{s},t-1},~\widehat{\alpha}_{\tn{b},t-1}$ and $\widehat{\f}_{t-1}$ into \eqref{eqn:opt-lmap-mat-a} to \eqref{eqn:opt-lmap-mat-z}.
\STATE $\bar{\bm\Upsilon}_t=\left(\D^\hr\D +\m^\hr\m+\gamma_{t-1}\I_K\right)^{-1}\left(\D^\hr\x+\m^\hr\z+\gamma_{t-1}\rt_{t-1}\right)$
\STATE $\bar{\gamma}_t=K\tn{Tr}\left(\left[\D^\hr\D + \m^\hr\m+\gamma_{t-1}\I_K\right]^{-1}\right)^{-1}$
\STATE $\widetilde{\gamma}_t=\bar{\gamma}_t-\gamma_{t-1}$
\STATE $\widetilde{\rt}_t=\widetilde{\gamma}_t^{-1}\left(\bar{\bm\Upsilon}_t\bar{\gamma}_t-\rt_{t-1}\gamma_{t-1}\right)$
\vspace{5pt}
\STATE // Separable MAP Projector
\STATE $\widehat{{\bm\Upsilon}}_t=\g_1\left(\widetilde{\rt}_t\right)$
\STATE $\widehat{\gamma}_t=\widetilde{\gamma}_t^{-1}\left\langle \g'_1\left(\widetilde{\rt}_t\right)\right\rangle$.
\STATE $\gamma_t=\widehat{\gamma}_t-\widetilde{\gamma}_t$
\STATE $\rt_t=\gamma_t^{-1}\left(\widehat{\gamma}_t\widehat{{\bm\Upsilon}}_t-\widetilde{\gamma}_t\widetilde{\rt}_t\right)$ \vspace{5pt}
\STATE //Find $\alpha?s,~\alpha?b$ and $\f$ through their closed-form solutions.
\STATE Compute $\widehat{\CC}_t,~\widehat{\K}_t$ and $\widehat{\e}_t$ by substituting $\widehat{{\bm\Upsilon}}?t$ into \eqref{eqn:precod-mat-c}, \eqref{eqn:precod-mat-k} and \eqref{eqn:precod-mat-e}.
\STATE $\widehat{\alpha}_{\tn{s},t}=g_4\left(\widehat{\CC}_t\right)$
\STATE $\widehat{\alpha}_{\tn{b},t}=g_5\left(\widehat{\K}_t,~\widehat{\e}_t\right)$
\STATE $\widehat{\f}?t=g_6\left(\widehat{\K}_t,~\widehat{\e}_t\right)$
\STATE $t \leftarrow t+1$ \vspace{5pt}
\UNTIL $|E_t-E_{t-1}|<\epsilon E_{t-1}$ or $t>T?{MAX}$
\RETURN $\widehat{{\bm\upsilon}}_t, \ \widehat{\f}_t, \ \widehat{\alpha}_t$.
\end{algorithmic}
\end{algorithm*}

Since we have now found the solution to both optimization problems in \eqref{eqn:ao-phase-mat} and \eqref{eqn:ao-precod-mat}, we can combine their solutions together into one algorithm. We define the MSE at iteration $t$ as follows:
\begin{equation}
E_t~\triangleq~f\big(\widehat{\alpha}_{\tn{s},t},~\widehat{\alpha}_{\tn{b},t},~\widehat{\f}_t,~\widehat{\bm\Upsilon}_t\big).
\end{equation}
The algorithm stops when $|E_t-E_{t-1}|<\epsilon E_{t-1}$, where $\epsilon \in \R_+$ is some precision tolerance. The overall block diagram and the algorithmic steps are shown, respectively, in Fig. \ref{fig:overall} and \textbf{Algorithm \ref{algo:overall}}. The convergence and complexity of the OOVAMP-based approach are discussed in \cite{haseebirs}. Because of the monotone convergence theorem in real analysis \cite{monotone}, \textbf{Algorithm \ref{algo:overall}} is guaranteed to converge since the MSE is minimized in every step and the objective function, $f(\alpha?s, \alpha?b, \f, {\mathbf\Upsilon})$, is lower bounded by zero. The algorithm can be efficiently implemented by exploiting matrix structures, and by using the singular value decomposition (SVD) form of OOVAMP so that the computational complexity is of $\BO(MNL(K+N))$.

\section{Numerical Results and Performance Analysis}
\label{sec:numeric-res}

\subsection{Simulation Model and Parameters}

We present Monte-Carlo simulation results to assess the performance of the proposed MIS-based communication scheme. We assume that the MIS is located at a fixed distance of $500$ m from the BS and the users are spread uniformly at a radial distance of $10$ m to $50$ m from the MIS. A path-based propagation channel model, also known as parametric channel model \cite{heathmimo}, is used. Such a model is more appropriate for systems with large antenna arrays. One key parameter of such a channel model is the number of multi-path components of the BS-MIS channel which governs the effect of channel correlation.
\begin{table}[!ht]
\centering
\caption{Simulation parameters with their notations and values}
\label{table:sim-parameter}
\begin{tabular}{|p{0.39\textwidth}|P{0.13\textwidth}|p{0.26\textwidth}|P{0.15\textwidth}|}
\hline 
\vspace{-10pt} \center{\textbf{Parameter}}  &  \vspace{-5pt}\textbf{Notation, Value}  & \vspace{-10pt} \center{\textbf{Parameter}}  &  \vspace{-5pt}\textbf{Notation, Value}\\
 \hline
\hline 
 Number of channel paths in the MIS-BS link & $Q?{MIS}=10$ & 
 MIS-BS distance & $d?{MIS}=500$ m\\
\hline
 Number of channel paths in the BS-user link & $Q?{b-u}=2$ &
 User-BS distance & $d=500$ m\\
\hline 
 Number of channel paths in the MIS-user link & $Q?{s-u}=2$&
 User-MIS distance & $d' \in [10,50]$ m \\
\hline 
 Path-loss exponent for the MIS-BS, MIS-user link & $\eta=2.5$&
 Noise variance & $\sigma?w^2=-100$ dBm \\
\hline 
 Path-loss exponent for the BS-user link & $\eta=3.7$&
 Channel path gains distribution & $c_q\sim\CG\N(0,1)$ \\
\hline 
 Reference distance & $d?0=1$ m &
 Path-loss at the reference distance & $C?0=-30$ dB \\
\hline
\end{tabular}
\end{table}

Assuming a uniform linear array with $N$ antennas at the BS and a square uniform planar array with $K$ antenna elements at the MIS, the channel between the MIS and the BS is generated according to:
\begin{equation}
\label{eqn:chanl-b-s}
\h?{b-s}~=~\sqrt{L(d?{MIS})}\ds\sum_{q=1}^{Q?{MIS}}\cn_q\as?{MIS}(\varphi_q,\psi_q)\as?{BS}(\phi_q)^\tr.
\end{equation}
The channel vectors for the link between each single antenna $m$-th user and the MIS, and each $m$-th user and the BS are  modeled, respectively, as follows:
\begin{align}
\label{eqn:chanl-s-u}
\hs_{\tn{s-u},m}&~=~\sqrt{L(d'_m)}\ds\sum_{q=1}^{Q?{s-u}}\cn_{m,q}\as?{MIS}(\varphi_{m,q},\psi_{m,q}), \quad m=1, \cdots, M,\\
\label{eqn:chanl-b-u}
\hs_{\tn{b-u},m}&~=~\sqrt{L(d_m)}\ds\sum_{q=1}^{Q?{b-u}}\cn_{m,q}\as?{BS}(\phi_{m,q}), \quad m=1, \cdots,M.
\end{align}

The term $L(d)=C?0(d/d?0)^{\eta}$ in \eqref{eqn:chanl-b-s}, \eqref{eqn:chanl-b-u}, \eqref{eqn:chanl-s-u} is the distance-dependent path-loss factor, where $C?0$ denotes the path-loss at a reference distance $d?0 = 1$ m, and $\eta$ is the path-loss exponent. To account for channel correlation effects, we set the number of multi-path components lower than the number of BS antennas and the MIS antenna elements for the BS-MIS channel, thereby making the channel matrix rank-deficient. Therefore, in the simulations, we set the number of BS-served users lower than the rank of the BS-MIS channel matrix $\h?{b-s}$. For brevity, the details on the array steering vectors, $\as?{MIS}$ and $\as?{BS}$, are omitted, and interested readers are referred to \cite{haseebirs} for complete information on the channel model. Table \ref{table:sim-parameter} describes the various simulation parameters along with their values used in the simulations. Moreover, we split the total transmit power, $P$, between the BS- and MIS-served users according to the share of each type of users, i.e., $P?b=\frac{B}{M}P$ and $P?s=\frac{R}{M}P$. The results are averaged over $1000$ independent Monte-Carlo trials.

We use the \textit{sum-rate}, $\widehat{C}$, for performance evaluation which is defined as follows:
\begin{equation}
\widehat{C}~=~\ds\sum_{m=1}^M\log_2\left(\frac{1}{\tn{MMSE}_m}\right),
\end{equation}
where $\tn{MMSE}_m$ refers to the MSE of each $m$-th user's received symbol.
Since the proposed approach of using the RIS as a modulating surface (i.e., MIS) is novel, we benchmark it against the following two traditional schemes  wherein the RIS is merely used for passive beamforming purposes in combination with MMSE precoding at the BS:
\begin{itemize}
\item[i.] \textsc{Scheme 1}: a multi-user  MIMO system assisted by one RIS where the optimization of the phase matrix  is solved through alternating optimization and OOVAMP.
\item[ii.] \textsc{Scheme 2}: a multi-user MIMO system assisted by one RIS where the optimization of the phase matrix is performed using the semidefinite relaxation (SDR) technique.
\end{itemize}

\subsection{Performance Results With Perfect CSI}
\label{sec:sub-perfect-csi}
\begin{figure}
\centering
\begin{subfigure}{.5\textwidth}
\centering
\includegraphics[scale=0.55]{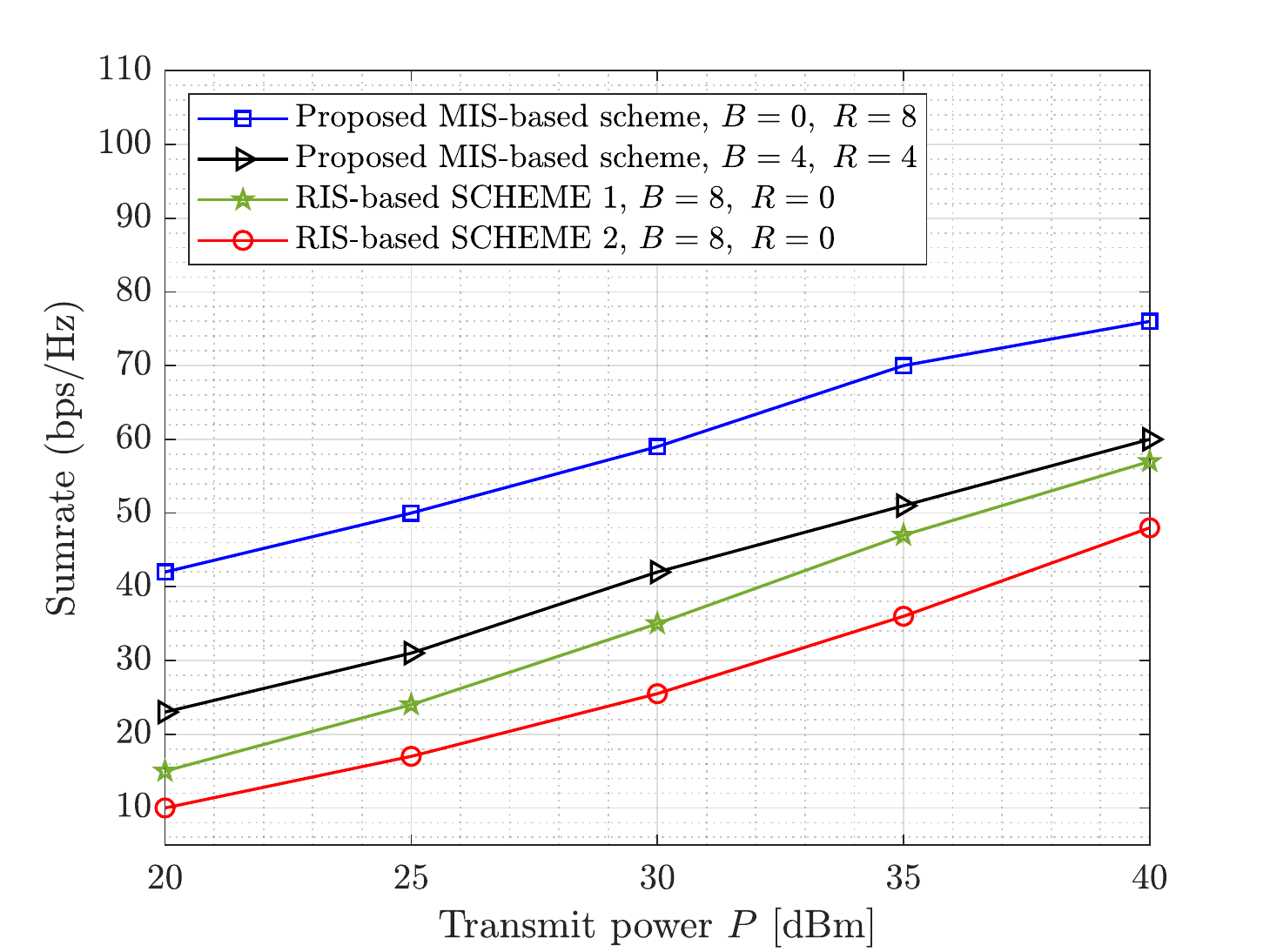}
\caption{\footnotesize $M=8,~N=32$, and\\ $K=256$.}
\label{fig:sum-ptr}
\end{subfigure}%
\begin{subfigure}{.5\textwidth}
\centering
\includegraphics[scale=0.55]{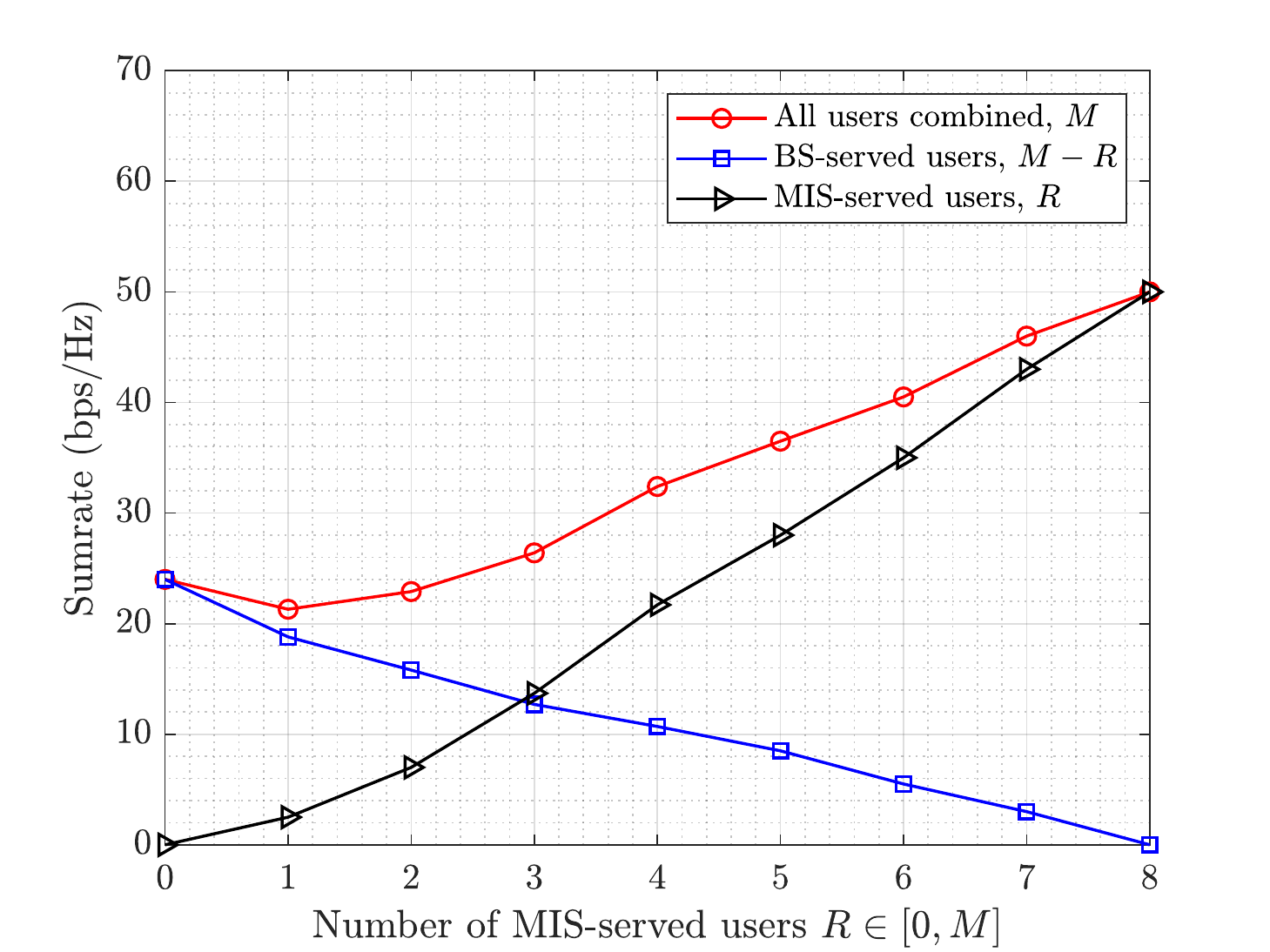}
\caption{\footnotesize $M=8, \ N=32$, and\\ $P=25$ dBm.}
\label{fig:sum-irs-users}
\end{subfigure}%
\caption{(a) Sum-rate versus transmit power, and (b) Sum-rate versus the number of MIS-served users.}
\end{figure}

We consider a typical urban or suburban environment where the BS is located faraway from the users and has no LOS to them. However, the MIS is installed at a location where an LOS component is present in the BS-MIS link but not in the user-MIS link. We also set the number of BS antennas to $N=32$ and the number of MIS reflective elements to $K=256$.
Fig. \ref{fig:sum-ptr}, depicts the achievable sum-rate versus the transmit power, $P$, for the different considered transmission schemes. The total number of users is equal to $M=8$.
Here we consider two configurations for the proposed approach: $i)$ when all the users are served by  the MIS, and $ii)$ a hybrid case wherein the MIS and the BS both serve $4$ users each. The configuration in which all the users are served by the MIS significantly outperforms the scheme in which the RIS is only used for beamforming  (i.e., \textsc{Scheme 1}). This is because, the MIS-served users do not suffer from the path-loss between the BS and the MIS. Moreover, at low transmit power (e.g., $P=20$ dBm), the sum-rate for the users solely served by the MIS is, respectively, two and four times the sum-rate of the BS-served users for the beamforming-only RIS-assisted OOVAMP-based and SDR-based approaches. For the hybrid configuration, the resulting sum-rate edges \textsc{Scheme 1} but, it is lower than the case when all users are served by the MIS. This confirms that it is more beneficial to serve the users by the MIS.

Fig. \ref{fig:sum-irs-users} shows a plot of the sum-rate against the share of MIS-served users among the total number of users. It is observed that the combined sum-rate first decreases and then monotonically increases with the number of MIS-served users. This is because of the presence of cross-user interference among the MIS-served and the BS-served users since the MIS is performing both tasks, i.e., beamforming to assist the BS and also data embedding to serve another set of users.
This implies that there is more loss than gain when the ratio of the MIS-served users to the total number of users becomes small. 

\begin{figure}
\centering
\begin{subfigure}{.5\textwidth}
\centering
\includegraphics[scale=0.55]{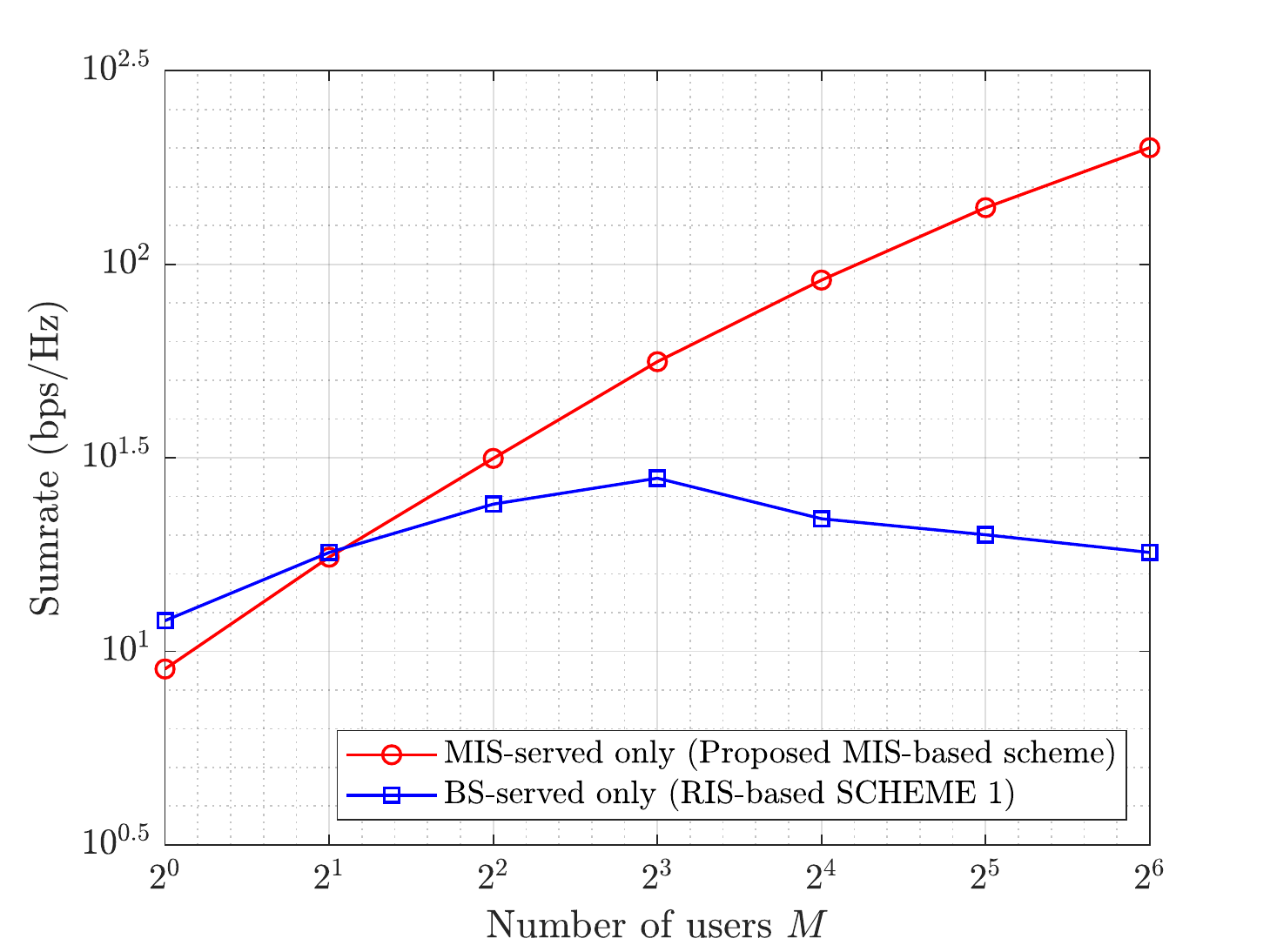}
\caption{\footnotesize $N=32,~K=256$ and\\ $P=30$ dBm.}
\label{fig:sum-users}
\end{subfigure}%
\begin{subfigure}{.5\textwidth}
\centering
\includegraphics[scale=0.55]{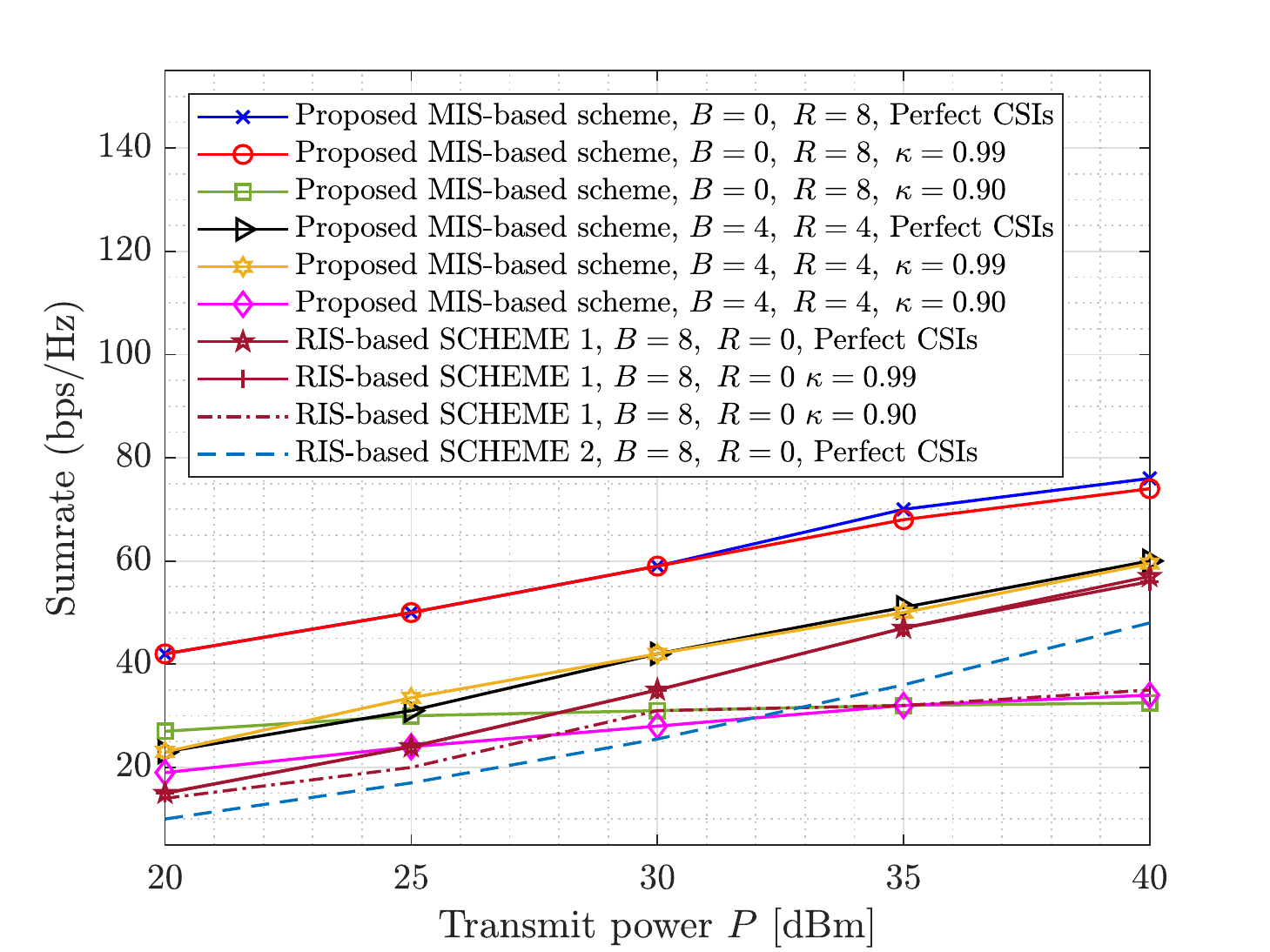}
\caption{\footnotesize $M=8,~N=32$, and\\ $K=256$.}
\label{fig:imperfect-csi}
\end{subfigure}
\caption{(a) Sum-rate versus the number of users, and (b) Sum-rate versus transmit power under imperfect CSI.}
\end{figure}

Fig. \ref{fig:sum-users} illustrates the sum-rate versus the total number of users being served. The users are solely served by the BS for one plot and by the MIS for the other. Here we show the benefits of the approach of using the MIS as a modulating surface to directly serve users. Although the sum-rate of the BS-served users is higher when the number of users becomes small, the sum-rate of the MIS-served users keeps increasing with the number of users while the sum-rate of the BS-served users only increases up to the number of available channel paths which is set to $10$. This is because, the number of users that can be served by the MIS is independent of the number of BS antennas and the correlation in the MIS-BS channel. The upper limit for the number of MIS-served users is equal to the number of MIS antenna elements $K$.

\subsection{Performance Results With Imperfect CSI}

In this section, we asses the performance of the proposed scheme in the presence of channel estimation errors. Specifically, we consider a scenario where pilot training followed by MMSE estimation algorithms are used to estimate the cascaded BS-MIS-users and the direct BS-users channels \cite{qauchannel1,qauchannel2}. We model the estimated channel matrix and vectors using the statistical CSI error model proposed in \cite{giffordcsi,annacsi,zhangcsi} as follows:
\begin{equation}
\widehat{\h}?{b-s}=\kappa\h?{b-s} + \sqrt{(1-\kappa^2)L(d?{MIS})}{\bm\Delta}?{b-s},
\end{equation}
\begin{equation}
\widehat{\hs}_{\tn{b-u},m}=\kappa\hs_{\tn{b-u},m} + \sqrt{(1-\kappa^2)L(d_m)}{\bm\delta}_{\tn{b-u},m}, \quad m=1, \cdots, M,
\end{equation}
\begin{equation}
\widehat{\hs}_{\tn{s-u},m}=\kappa\hs_{\tn{s-u},m} + \sqrt{(1-\kappa^2)L(d'_m)}{\bm\delta}_{\tn{s-u},m}, \quad m=1, \cdots, M,
\end{equation}
where $\kappa \in [0,1]$ denotes the channel estimation accuracy and ${\bm\Delta}?{b-s}$, ${\bm\delta}_{\tn{b-u},m}$ and ${\bm\delta}_{\tn{s-u},m}$ follow the circularly symmetric complex Gaussian (CSCG) distribution, i.e.,
\mbox{$\tn{vec}({\bm\Delta}?{b-s}) \sim \CG\N(\0, \1_{N \times N} \otimes \I_K)$,} ${\bm\delta}_{\tn{b-u},m} \sim \CG\N(\0, \I_N)$ and ${\bm\delta}_{\tn{s-u},m} \sim \CG\N(\0, \I_K)$. We first optimize the variables $\alpha?s,~\alpha?b,~\f$ and ${\bm\Upsilon}$ under imperfect CSI and then use the exact CSI matrices to calculate the sum rate.

Fig. \ref{fig:imperfect-csi} plots the sum-rate versus transmit power for different values of the channel estimation accuracy parameter $\kappa$. We also include plots for the other schemes (i.e., \textsc{Schemes 1 and 2}) which use the RIS as a purely reflective surface) under perfect CSI for reference. The results show the resilience of the proposed MIS-based approach against small channel estimation errors. At low SNR, it is observed that the proposed design with a low channel estimation accuracy of $\kappa=0.90$ performs better than the baseline schemes even under perfect CSIs. Moreover, the performance loss is negligible with a high channel estimation accuracy value of $\kappa=0.99$.

\section{Conclusion}
\label{sec:conclusion}
We have presented a novel approach of employing the MIS for performing passive beamforming and data embedding for the BS-served and the MIS-served users, respectively, in a multi-user MIS-assisted downlink MIMO system. The associated joint convex optimization problem has been formulated under the sum MMSE criterion in order to maximize the users' spectral efficiency. Alternating minimization has been used to split the original optimization problem into two tasks, i.e., separately optimizing the MIS phase shifts and jointly optimizing the BS precoding and the receive scaling factors for the BS- and MIS-served users. The optimal solution to the  joint optimization problem for precoding is found in closed form. We have optimized the MIS phase shifts using OOVAMP by deriving the problem-specific modules of the OOVAMP algorithm. Moreover, the original joint problem has been solved under both the ideal and a practical constraint on the MIS phase shifts. Simulation results illustrate highly superior system throughput performance of the proposed MIS-based scheme over two baseline schemes in which the MIS is used for beamforming purposes only. Moreover, the proposed approach can support more number of users simultaneously than existing beamforming approaches wherein the users are served by the BS only. Finally, the results under imperfect CSI confirm that the performance remains nearly unchanged in the presence of small channel estimation errors.

Three possible extensions to the proposed framework for MIS-assisted networks are outlined as follows:
\begin{itemize}
    \item \textit{Optimal power allocation:} In the present work, the power split between the BS- and MIS-served users was not optimized. In fact, we simply split the power between the BS- and MIS-served users according to the share of each type of users. To utilize the transmit power efficiently, the joint optimization problem can be extended to optimally allocate power between the BS- and MIS-served users.
    \item \textit{Optimal user assignment:} Instead of splitting the users arbitrarily between the BS and the MIS, the decision must be taken on the basis of CSI. The users for which the direct BS-user link is strong enough to provide higher data rate than if they were served by the MIS,  can be served and beamformed by the BS and the MIS, respectively. Similarly, the  users that have a weak direct BS-user link can be served through modulation by the MIS. User assignment can be optimized to maximize system throughput.
    \item \textit{Jointly optimized reference carrier signal:} Presently, we assign the strongest path in the BS-MIS link to transmit the reference carrier signal pertaining to the MIS-served users. The remaining paths are used to serve the BS-served users which results in lower sum-rate for BS-served users. To mitigate this problem, the reference transmitted vector, $\vs?b$, can also be jointly optimized with the other variables to enhance the overall throughput of the system.
\end{itemize}
\appendices
\section{}
\label{apd:A}
\noindent
We have the following function:
\begin{equation}
\label{eqn:apda-exp}
   f(\alpha?s, \alpha?b, \f, {\mathbf\Upsilon})= \E_{\yn,\sn}\left\lbrace\norm{\yn-\sn}?F^2\right\rbrace,
\end{equation}
where
\begin{align}
\begin{split}
        \yn&=\alpha?b\Big[\h?{s-u}^\hr\tn{Diag}(\bm\upsilon_1)\h?{b-s}\f\ssn_{\tn{b},1}+ \h?{b-u}^\hr\f\ssn_{\tn{b},1},\cdots, \h?{s-u}^\hr\tn{Diag}(\bm\upsilon_L)\h?{b-s}\f\ssn_{\tn{b},L}+ \h?{b-u}^\hr\f\ssn_{\tn{b},L}\Big]\\
        &+\alpha?s\sqrt{P?s}\Big[\h?{s-u}^\hr\tn{Diag}(\bm\upsilon_1)\h?{b-s}\vs?b, \cdots, \h?{s-u}^\hr\tn{Diag}(\bm\upsilon_L)\h?{b-s}\vs?b\Big]\\
        &+\Big[\alpha?b\wn?b^\tr, ~ \alpha?s\wn?s^\tr\Big]^\tr,
\end{split}\\
\sn&=\big[\sn?b^\tr,~\s?s^\tr\big]^\tr.
\end{align}
We explicitly write the function in \eqref{eqn:apda-exp} and then expand it as follows:
\begin{align}
    f(\alpha?s, \alpha?b, \f, {\mathbf\Upsilon}) &=\E_{\yn,\sn}\left\lbrace\tn{Tr}\left(\yn^\hr\yn-\yn^\hr\sn-\sn^\hr\yn+ \sn^\hr\sn\right)\right\rbrace\\
    \begin{split}
         &=\E_{\yn,\sn?b}\Big\lbrace\tn{Tr}\Big(\yn^\hr\yn- \yn^\hr\big[\sn?b^\tr,~\0_{L \times R}\big]^\tr-
         \yn^\hr\big[\0_{L \times B},~\s?s^\tr\big]^\tr
         -\big[\sn?b^\hr,~\0_{L \times R}\big]\yn\\
         &-\big[\0_{L \times B},~\s?s^\hr\big]\yn+
         \big[\sn?b^\hr,~\0_{L \times R}\big]\big[\sn?b^\tr,~\0_{L \times R}\big]^\tr
         +\big[\sn?b^\hr,~\0_{L \times R}\big]\big[\0_{L \times B},~\s?s^\tr\big]^\tr\\
         &+\big[\0_{L \times B},~\s?s^\hr\big]\big[\sn?b^\tr,~\0_{L \times R}\big]^\tr+
         \big[\0_{L \times B},~\s?s^\hr\big]\big[\0_{L \times B},~\s?s^\tr\big]^\tr
         \Big)\Big\rbrace \notag
    \end{split}\\
    \label{eqn:apda-exp-terms}
    \begin{split}
        f(\alpha?s, \alpha?b, \f, {\mathbf\Upsilon}) &=\E_{\yn,\sn?b}\Big\lbrace\tn{Tr}\Big(\yn^\hr\yn- \yn^\hr\big[\sn?b^\tr,~\0_{L \times R}\big]^\tr
         -\big[\sn?b^\hr,~\0_{L \times R}\big]\yn\\
         &-\yn^\hr\big[\0_{L \times B},~\s?s^\tr\big]^\tr
         -\big[\0_{L \times B},~\s?s^\hr\big]\yn
         +\big[\sn?b^\hr,~\0_{L \times R}\big]\big[\sn?b^\tr,~\0_{L \times R}\big]^\tr\\
         &+\big[\0_{L \times B},~\s?s^\hr\big]\big[\0_{L \times B},~\s?s^\tr\big]^\tr
         \Big)\Big\rbrace.
    \end{split}
\end{align}
By defining the matrices:
\begin{align}
    \yn?b&=\alpha?b\Big[\h?{s-u}^\hr\tn{Diag}(\bm\upsilon_1)\h?{b-s}\f\ssn_{\tn{b},1}+ \h?{b-u}^\hr\f\ssn_{\tn{b},1},\cdots, \h?{s-u}^\hr\tn{Diag}(\bm\upsilon_L)\h?{b-s}\f\ssn_{\tn{b},L}+ \h?{b-u}^\hr\f\ssn_{\tn{b},L}\Big],\\
    \y?e&=\alpha?b\Big[\h?{s-u}^\hr\tn{Diag}(\bm\upsilon_1)\h?{b-s}\f+ \h?{b-u}^\hr\f,\cdots, \h?{s-u}^\hr\tn{Diag}(\bm\upsilon_L)\h?{b-s}\f+ \h?{b-u}^\hr\f\Big],\\
    \y?s&=\alpha?s\sqrt{P?s}\Big[\h?{s-u}^\hr\tn{Diag}(\bm\upsilon_1)\h?{b-s}\vs?b, \cdots, \h?{s-u}^\hr\tn{Diag}(\bm\upsilon_L)\h?{b-s}\vs?b\Big],\\
    \wn&=\Big[\alpha?b\wn?b^\tr, ~ \alpha?s\wn?s^\tr\Big]^\tr,
\end{align}
we expand the terms in \eqref{eqn:apda-exp-terms} and take expectation w.r.t. the random matrices $\sn?b$ and $\wn$ thereby leading to:

\begin{align}
\label{eqn:apda-term1}
    \begin{split}
      \E_{\yn,\sn?b}\big\lbrace\tn{Tr}\big(\yn^\hr\yn\big)\big\rbrace
      &=\E_{\wn, \sn?b}\big\lbrace\tn{Tr}\big(\yn?b^\hr\yn?b + \yn?b^\hr\y?s +\yn?b^\hr\wn  + \y?s^\hr\yn?b + \y?s^\hr\y?s \\
      &+\y?s^\hr\wn +\wn^\hr\yn?b +\wn^\hr\y?s + \wn^\hr\wn\big)\big\rbrace
    \end{split}\\
      &=\tn{Tr}\big(\y?e^\hr\y?e +\y?s^\hr\y?s\big)+LB\sigma?w^2\alpha?b^2 + LR\sigma?w^2\alpha?s^2, \notag\\
\label{eqn:apda-term2}
    \begin{split}
      \E_{\yn,\sn?b}\Big\lbrace\tn{Tr}\Big(\big[\sn?b^\hr,~\0_{L \times R}\big]\yn\Big)\Big\rbrace 
      &=\E_{\wn, \sn?b}\Big\lbrace\tn{Tr}\Big(\big[\sn?b^\hr,~\0_{L \times R}\big]\yn?b + \big[\sn?b^\hr,~\0_{L \times R}\big]\y?s\\
      &+ \big[\sn?b^\hr,~\0_{L \times R}\big]\wn\Big)\Big\rbrace
    \end{split}\\
     &=\tn{Tr}\bigg(\Big[[\I_{B,1}, \ \0_{B\times R,1}]^\tr, \cdots, [\I_{B,L}, \ \0_{B\times R,L}]^\tr \Big]^\hr\y?e\bigg), \notag\\
\label{eqn:apda-term3}
    \begin{split}
        \E_{\yn,\sn?b}\Big\lbrace\tn{Tr}\Big(\big[\0_{L \times B},~\s?s^\hr\big]\yn\Big)\Big\rbrace
        &=\E_{\wn, \sn?b}\Big\lbrace\tn{Tr}\Big(\big[\0_{L \times B},~\s?s^\hr\big]\yn?b + \big[\0_{L \times B},~\s?s^\hr\big]\y?s\\
        &+ \big[\0_{L \times B},~\s?s^\hr\big]\wn\Big)\Big\rbrace
    \end{split}\\
     &=\tn{Tr}\Big(\big[\0_{L \times B},~\s?s^\hr\big]\y?s\Big), \notag
\end{align}
and
\begin{align}
\label{eqn:apda-term4}
         \begin{split}
         \E_{\sn?b}\bigg\lbrace\tn{Tr}\Big(\big[\sn?b^\hr,~\0_{L \times R}\big]\big[\sn?b^\tr,~\0_{L \times R}\big]^\tr \Big)\bigg\rbrace
         &=\tn{Tr}\bigg(\Big[[\I_{B,1}, \ \0_{B\times R,1}]^\tr, \cdots, [\I_{B,L}, \ \0_{B\times R,L}]^\tr \Big]^\hr\\
         &\times \Big[[\I_{B,1}, \ \0_{B\times R,1}]^\tr, \cdots, [\I_{B,L}, \ \0_{B\times R,L}]^\tr \Big]\bigg).
     \end{split}
\end{align}
The remaining two terms in \eqref{eqn:apda-exp-terms} can be computed by following \eqref{eqn:apda-term2} and \eqref{eqn:apda-term3}. Finally, by substituting \eqref{eqn:apda-term1}$-$\eqref{eqn:apda-term4} into \eqref{eqn:apda-exp-terms} and expressing the function in the form of norms we obtain:
\begin{align}
    \begin{split}
    f(\alpha?s, \alpha?b, \f, {\mathbf\Upsilon})
    &=\bigg\|\alpha?b\Big[\h?{s-u}^\hr\tn{Diag}(\bm\upsilon_1)\h?{b-s}\f+ \h?{b-u}^\hr\f,\cdots, \h?{s-u}^\hr\tn{Diag}(\bm\upsilon_L)\h?{b-s}\f+ \h?{b-u}^\hr\f\Big]\\
    &-\Big[[\I_{B,1}, \ \0_{B\times R,1}]^\tr, \cdots, [\I_{B,L}, \ \0_{B\times R,L}]^\tr \Big]\bigg\|?F^2\\
    &+\bigg\|\alpha?s\sqrt{P?s}\Big[\h?{s-u}^\hr\tn{Diag}(\bm\upsilon_1)\h?{b-s}\vs?b, \cdots, \h?{s-u}^\hr\tn{Diag}(\bm\upsilon_L)\h?{b-s}\vs?b\Big]-\left[\0_{L \times B} \ \s?s^\tr\right]^\tr\bigg\|?F^2\\
    &+LB\sigma?w^2\alpha?b^2 + LR\sigma?w^2\alpha?s^2.
\end{split}
\end{align}

\section{}
\label{apd:B}
\noindent
We have the following two norms:
\begin{align}
\label{eqn:norm-bs}
\begin{split}
    f_1&~=~\bigg\|\alpha?b\Big[\h?{s-u}^\hr\tn{Diag}(\bm\upsilon_1)\h?{b-s}\f+ \h?{b-u}^\hr\f,\cdots, \h?{s-u}^\hr\tn{Diag}(\bm\upsilon_L)\h?{b-s}\f+ \h?{b-u}^\hr\f\Big]\\
    &~-~\Big[[\I_B, \ \0_{B\times R,1}]^\tr, \cdots, [\I_B, \ \0_{B\times R,L}]^\tr \Big]\bigg\|?F^2,
\end{split}\\
\label{eqn:norm-irs}
    f_2&~=~\bigg\|\alpha?s\sqrt{P?s}\Big[\h?{s-u}^\hr\tn{Diag}(\bm\upsilon_1)\h?{b-s}\vs?b, \cdots, \h?{s-u}^\hr\tn{Diag}(\bm\upsilon_L)\h?{b-s}\vs?b\Big]-\left[\0_{L \times B} \ \s?s^\tr\right]^\tr\bigg\|?F^2.
\end{align}
By defining the matrices, $\A=\alpha?b\h?{s-u}^\hr \in \C^{M \times K}$ and $\B =(\h?{b-s}\f)^\tr \in \C^{B \times K}$, we rewrite \eqref{eqn:norm-bs} as follows:
\begin{align}
\label{eqn:norm-bs-only}
\begin{split}
        f_1&~=~\bigg\|\Big[\A\tn{Diag}(\bm\upsilon_1)\B^\tr,\cdots, \A\tn{Diag}(\bm\upsilon_L)\B^\tr \Big]\\
        &~-~\Big[[\I_B, \ \0_{B\times R,1}]^\tr - \alpha?b\h?{b-u}^\hr\f, \cdots, [\I_B, \ \0_{B\times R,L}]^\tr - \alpha?b\h?{b-u}^\hr\f \Big]\bigg\|?F^2.
\end{split}
\end{align}
We then define a column-wise Khatri-Rao matrix, $\D \in \C^{MB \times K}$ and another matrix, $\x \in \C^{MB \times L}$, as follows:
\begin{align}
    \D &=\left[\bs_1 \otimes \as_1, \cdots, \bs_K \otimes \as_K\right],\\
    \x &=\big[\tn{vec}\left([\I_B, \ \0_{B\times R,1}]^\tr-\alpha?b\h?{b-u}^\hr\f\right), \cdots, \tn{vec}\left([\I_B, \ \0_{B\times R,L}]^\tr-\alpha?b\h?{b-u}^\hr\f\right)\big].
\end{align}
Through vectorization, we have the following relation for the norm of a matrix:
\begin{equation}
\label{eqn:norm-relation}
    \norm{\D\bm\upsilon-\tn{vec}(\CC)}?2^2=\norm{\A\tn{Diag}(\bm\upsilon)\B^\tr-\CC}?F^2.
\end{equation}
By using the relation in \eqref{eqn:norm-relation} and substituting the matrices $\D$ and $\x$ into \eqref{eqn:norm-bs-only}, we get:
\begin{equation}
    f_1=\norm{\D[\bm\upsilon_1,~\bm\upsilon_2, \cdots, \bm\upsilon_L]-\x}?F^2,
\end{equation}
or equivalently:
\begin{equation}
    f_1=\norm{\D\bm\Upsilon-\x}?F^2.
\end{equation}
Similarly, by defining the matrices:
\begin{align}
    \m &=\sqrt{P?s}\alpha?s\h?{s-u}^\hr\tn{Diag}(\h?{b-s}\vs?b),\\
    \z &=\left[\0_{L \times B}, \ \s?s^\tr\right]^\tr,
\end{align}
and then substituting them into \eqref{eqn:norm-irs} one obtains:
\begin{equation}
    f_2=\norm{\m\bm\Upsilon-\z}?F^2.
\end{equation}
\bibliographystyle{IEEEtran}
\bibliography{IEEEabrv,references} 
\end{document}